\documentclass[10pt,journal,compsoc]{IEEEtran}
\IEEEoverridecommandlockouts
\usepackage{cite}
\usepackage{amsmath,amsthm,amssymb,amsfonts,bm}

\newtheorem{lemma}{Lemma}

\newtheorem{prop}{Proposition}

\newtheorem{corol}{Corollary}

\newtheorem{obs}{Observation}

\newtheorem{question}{Question}
\newtheorem{defn}{Definition}
\usepackage{algorithmic}
\usepackage{graphicx}  
\usepackage{epstopdf}
\usepackage{textcomp}
\usepackage{xcolor}
\usepackage{url}

\usepackage{enumitem}
\usepackage{url}
\usepackage{float}
\setlist[itemize]{leftmargin=*}
\setlist[enumerate]{leftmargin=*}
\usepackage{multirow}
\usepackage{bbm}
\usepackage{subfigure}
\usepackage[linesnumbered, ruled]{algorithm2e}

\def\BibTeX{{\rm B\kern-.05em{\sc i\kern-.025em b}\kern-.08em
    T\kern-.1667em\lower.7ex\hbox{E}\kern-.125emX}}
\title{Location Privacy Protection Game against Adversary through Multi-user Cooperative Obfuscation }

\author{Shu Hong,
~\IEEEmembership{Graduate Student Member,~IEEE,}
and~Lingjie~Duan,~\IEEEmembership{Senior Member,~IEEE,}
\IEEEcompsocitemizethanks{\IEEEcompsocthanksitem 
S. Hong and L. Duan are with the Pillar of Engineering Systems and Design, Singapore University of Technology and Design, Singapore (E-mail: shu\_hong@mymail.sutd.edu.sg; lingjie\_duan@sutd.edu.sg).\protect\\

\IEEEcompsocthanksitem
Part of this work was presented at IEEE ISIT 2022 \cite{hong2022multi}.
}%

}

\begin{document}

\setlength{\abovedisplayskip}{2pt}
\setlength{\belowdisplayskip}{2pt}

\IEEEtitleabstractindextext{%
\begin{abstract}
In location-based services(LBSs), it is promising for users to crowdsource and share their Point-of-Interest(PoI) information with each other in a common cache to reduce query frequency and preserve location privacy. 
Yet most studies on multi-user privacy preservation overlook the opportunity of leveraging their service flexibility. 
This paper is the first to study multiple users' strategic cooperation against an adversary's optimal inference attack, by leveraging mutual service flexibility.
We formulate the multi-user privacy cooperation against the adversary as a max-min adversarial game and solve it in a linear program. 
Unlike the vast literature, even if a user finds the cached information useful, we prove it beneficial 
to still query the platform to further confuse the adversary.
As the linear program's computational complexity still increases superlinearly with the number of users' possible locations, we propose a binary obfuscation scheme in two opposite spatial directions to achieve guaranteed performance 
with only constant complexity. 
Perhaps surprisingly, 
a user with a greater service flexibility should query with a less obfuscated location to add confusion. 
Finally, we provide guidance  on the optimal query sequence among LBS users. Simulation results show that 
our crowdsourced privacy protection scheme greatly improves users' privacy as compared with existing approaches. 
\end{abstract}

\begin{IEEEkeywords}
Decentralized privacy preservation,
multi-user cooperative crowdsourcing,
location-based services with flexibility,
max-min adversarial game theory.
\end{IEEEkeywords}}

\maketitle

\IEEEpeerreviewmaketitle

\IEEEraisesectionheading{\section{Introduction}\label{sec:introduction}}
\IEEEPARstart{L}{ocation-based} services (LBSs) offer mobile users customized service recommendations with an integration of mobile users' geographic locations\cite{huang2022location}.  
To provide useful information about points of interests (PoIs) nearby (e.g., nightclubs and restaurants), the LBS platform responds to users' queries using their current locations.
%
Despite the customized service benefits, the usage of LBSs might leak users' private location information, as query data stored in LBS platforms may be revealed to advertisers or hacked by malicious attackers\cite{shaham2020privacy}.
%

To preserve users’ location privacy, both centralized and decentralized approaches are proposed and studied \cite{jiang2021location}.
The basic idea of centralized approaches is to introduce a trusted third party (TTP), which protects users’ privacy by operating between users and the LBS platform as the anonymizer (e.g., \cite{zhang2018dual,xiao2018cenlocshare,wang2018privacy,zhao2019synthesizing}). 
Such a TTP 
collects users' original queries and transmits the processed queries to the LBS platform after applying privacy-preserving techniques (e.g., mix zone, pseudonym).
Thus the LBS platform cannot identify the users' real locations.
%
However, a single-point failure at the TTP may lead to full privacy leakage of a large group of users \cite{newsofdatabreaches}.
A recent example is the leakage of user credentials from Okta, a third-party company that handles  log-ins for more than 100 million users \cite{Okta}. 

Decentralized approaches (e.g., \cite{qi2017privacy,xu2018check,gu2019privacy}) no longer rely on a TTP between the mobile users and the LBS platform.
There, a distributed user needs to initiate the LBS query by applying privacy-preserving techniques himself (e.g., cloaking \cite{li2021priparkrec}, dummy generation \cite{huang2021privacy}, $k$-anonymity \cite{buccafurri2021distributed} \cite{domingo2022decentralized} and differential privacy \cite{dwork2014algorithmic}\cite{jiang2021differential}).
However, such approaches may incur high overhead on the individual.
For example, the $k$-anonymity technique expects a user to obfuscate his real location with $k-1$ dummy locations, 
leading to a high complexity for computation and implementation at the end device.

Recently, some simple kinds of decentralized privacy-preserving approaches are proposed by using caching (e.g.,
\cite{zhang2019caching,shokri_hiding_2014,peng2017collaborative,hu2018proactive,jung2017collaborative,cui2020cache}). 
In \cite{zhang2019caching}, a user caches his prior query data to answer similar queries in the future and reduce the chances to leak his privacy to the untrusted or compromised LBS platform. 
In practice, however, an individual user's caching is far from enough to cover many PoIs for him to visit later. 
Thanks to crowdsourcing, it is more efficient for many users to share their queried PoIs with each other in a common cache.
Shokri \emph{et al.} in \cite{shokri_hiding_2014} proposed a user-collaborative privacy-preserving approach: once a user finds the formerly cached PoI information by other users helpful, he will no longer query the LBS platform.
Only if the user finds the shared PoI information in the cache 
not useful, he has to query the LBS  platform using his real location, which can be overheard by the adversary.   
Provided with users' overlapped mobility patterns and similar LBS interests, such cooperative approach efficiently reduces the overall query frequency for all users. 
If we relax to allow a user to make multiple queries and bear extra computational complexity or communication overhead, there are some other fusion works further combining caching with $k$-anonymity \cite{peng2017collaborative,hu2018proactive,jung2017collaborative} or $l$-diversity \cite{cui2020cache}.

It should be noted that once a user finds useful information in the cache, all the existing caching-based privacy protection approaches (\cite{zhang2019caching,shokri_hiding_2014,peng2017collaborative,hu2018proactive,jung2017collaborative,cui2020cache}) simply ask him to hide from the LBS platform without any query. We wonder if such hiding is beneficial to users, and this motivates the first key question of this paper.
\begin{question}
If a user already finds the shared PoI information useful in the cache, is it beneficial for him to hide from or further query the LBS platform? 
\end{question}

Actually, hiding from the LBS platform without query also reveals that the user's location is already covered in the existing cache, and we will prove in the paper that always querying helps confuse the adversary's optimal inference attack.

On the other hand, if the user does not find useful PoI information in the cache, we wonder how to add obfuscation to his query to still protect his privacy. 
\begin{question}
If a user finds the shared PoI information not useful in the cache, how to strategically add more obfuscation to his query?  
\end{question}

In the real world, many LBS users are actually flexible in service requirements and only expect that the returned PoIs are within a certain distance (e.g., restaurants and hotels within 1 km) \cite{urban2017commerce,LBSreport}.
An LBS user may query with an obfuscated location instead of his real location to leverage his service flexibility \cite{gursoy2018utility}. 
Some work further models this flexibility as a service requirement constraint \cite{shokri2016privacy}.
Yet most studies on
multi-user privacy preservation overlook the opportunity of leveraging service flexibility. 
To our best knowledge, this paper is the first to leverage service flexibility in multi-user privacy protection.

On the other hand, the LBS platform or any other cooperative peers may be compromised by the adversary to leak the shared information in the cache and should be regarded as untrusted.
The adversary is also aware of the users' service flexibility to add obfuscation to their queries, and may adaptively change the inference attack strategy. To proactively design the multi-user privacy protection mechanism, we should be first prepared to understand the adversary's best inference attack. 
This leads to the third key question of this paper:
\begin{question}
What is the adversary's best inference attack to the multi-user strategic LBS querying?    
\end{question}

It is natural to use game theory to model the interaction between the users' strategic queries and the adversary's inference attack. 
The adversary might hack into other untrusted peers or the platform for reported locations in the cache, yet it does not know users' private locations. Then we will model the interaction between users and the adversary as a max-min adversarial Bayesian game 
and accordingly design users' robust query strategies.

The key novelty and the main results of the paper are summarized as follows.

\begin{itemize}
\item 
\emph{Multi-user privacy cooperation by Leveraging service flexibility:}
To the best of our knowledge, this is the first paper to study how multiple users cooperate to query with maximum obfuscation against the adversary’s optimal inference attack, by leveraging their mutual service flexibility. We instruct users to not only share their searched PoI information in a common cache to reduce overall query frequency, but also cooperate to add maximum obfuscation to their queries in LBS. We consider the robust defence against an intelligent adversary, who knows the cached data by former users and the users' objective functions to reverse-engineer and infer users' locations from their queries.

\item 
\emph{Adversarial Bayesian game against optimal inference attack:}
As the users' locations are private information to be inferred by the adversary, we naturally formulate multi-user privacy cooperation against the adversary as a $\max$-$\min$ adversarial Bayesian game. We manage to simplify it to a linear programming (LP) problem, yet its computational complexity still increases superlinearly with the number of users’ possible locations.
We prove it beneficial for users to always query the LBS platform to add maximum obfuscation, even if they already find useful PoI information in the crowdsourced cache.

\item 
\emph{Approximate obfuscation cooperation schemes with low complexity:}
To greatly save the complexity and derive the closed-form defence solution, we propose a binary approximate obfuscation scheme with only constant complexity for users located on a one-dimensional (1D) line interval (e.g., avenue or road).
Depending on whether a user finds the shared PoI information useful or not, this approximation scheme tells how to misreport his query randomly in two opposite spatial directions. This scheme is easy to implement and we also extend it to users located in the two-dimensional (2D) plane to apply randomized misreporting in four spatial directions. 

\item 
\emph{Guaranteed multi-user privacy gain:}
Our binary approximate obfuscation scheme is proved to guarantee at least 3/5 of the optimal privacy gain. 
Perhaps surprisingly, we choose to instruct a user with a greater service flexibility  to query with a less obfuscated location to strategically confuse the adversary. 
We also prove the asymptotic optimum of our approximate obfuscation scheme, as long as there are enough number of crowdsourcing users.
Extensive simulations show our scheme significantly outperforms the state-of-the-art schemes.

\item 
\emph{Guidelines for the multi-user query sequence:}
Besides guiding each user's LBS query location, we further enhance the multi-user privacy protection performance, by optimizing the query sequence of users with different service flexibilities.
For the case of two users of similar small service flexibilities to cooperate, we prove it beneficial for the user with less service flexibility to query and help preserve the other user's privacy. Yet the sequence should reverse if they have very diverse service flexibilities. We also simulate the more general multi-user case to show similar insights for the optimal query sequence. 

\end{itemize}

The outline of the paper is organized as follows. 
Section \ref{Sec:model} presents the system model under the multi-user privacy preservation.
Section \ref{Sec: pri coop against ad inference} formulates each user's privacy cooperation problem against the adversary.
Section \ref{Sec:approximation1D} studies a approximate obfuscated query scheme in closed-form in a one-dimensional line. 
Section \ref{Sec:performance eva} evaluates the performance of the approximate scheme. 
Section \ref{Sec:order} studies the optimal query sequence to maximize the total expected privacy gain.
Section \ref{Subsec:extension 2D} extends the approximate scheme to the two-dimensional scenario.
Finally, Section \ref{Sec:conclusion} concludes this paper.
Due to space limit, we put the detailed proofs in the supplementary document.

\section{System Model and Problem Formulation}
\label{Sec:model}

\begin{figure}[!t]
\centering
\includegraphics[width=2.4 in]{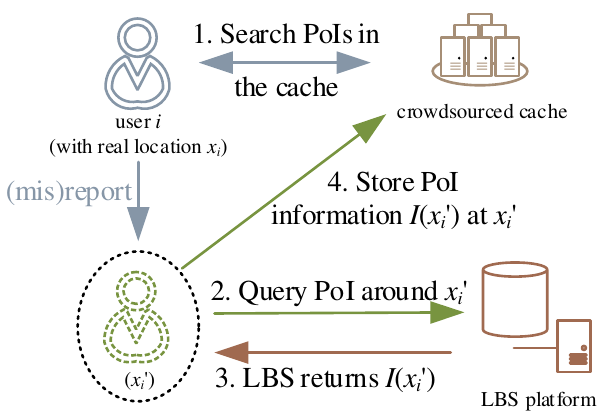}
\caption{Users' crowdsourced query scheme in LBS: user $i$ with real location $x_i$ examines the crowdsourced cache first, finding useful/unuseful PoI information. Then he may query the LBS platform with an obfuscated location $x_i'$ and receive the PoI information $I(x_i')$ at location $x_i'$. To benefit latter users, he stores $(x_i',I(x_i'))$ to share in the cache. 
}
\label{Fig:Two users' cooperative reportings}
\end{figure}

We consider $N$ active LBS users in a set $\mathcal{N}=\{1,2,\cdots,N\}$ with real locations $x_1,x_2,\cdots,x_N$ known to themselves only. They need nearby PoI information in a continuous bounded location set $\mathcal{M}$, which can be either a one-dimensional (1D) avenue line or a two-dimensional (2D) ground plane.
To learn useful PoI information from each other, a crowdsourced cache is used to enable a user to post and share his searched PoI information with latter users with similar PoI interests. 
To best use the cache, it is the best for users to sequentially demand PoI information from the LBS such that the latter can learn from former users' queries. 

Next, we first introduce the multi-user crowdsourced query scheme with obfuscation, then introduce users' randomized strategies against the adversary's inference attack in this scheme. Finally, we model the strategic interaction between cooperative users and the adversary as an adversarial Bayesian game.

\subsection{Multi-user Crowdsourced Query Scheme with Obfuscation}
Without loss of generality, we suppose user $i$ is the $i$-th among $N$ users to query.
Later in Section \ref{Sec:order}, besides studying each user's query strategy, we will further study the optimal query sequence among $N$ users from the perspective of the crowdsourcing system to maximize the users' total expected privacy. 

Each user $i\in \mathcal{U}$ with real location $x_i \in \mathcal{M}$ queries the LBS platform for nearby PoI information (e.g., hotels), by providing a location $x_i'\in \mathcal{M}$. The LBS platform will return PoI information $I(x_i')$ at $x_i'$ to the user.
%
In practice, the user is flexible to demand PoIs, and finds PoIs useful as long as they are within a certain distance $Q_i$ from his real location $x_i$ (e.g., \cite{li2009tradeoff,urban2017commerce}).
Formally, we give the definition of a user's service flexibility in the following.
\begin{defn}[User $i$'s service flexibility]
When searching for PoIs in LBSs, user $i$ with real location $x_i$ is flexible to accept the returned PoI information $I(x_i')$ at $x_i'$ if $D(x_i,x_i') \leq Q_i$, where $D(\cdot, \cdot)$ measures the Euclidean distance between any two points, the parameter $Q_i$ measures the user's service flexibility.
\end{defn}

One can imagine the scenario when an LBS user searches for nearby restaurant recommendations, usually he does not require the returned PoI information to be exactly at his real location, given he is flexible to walk or drive a certain distance.
The user requirements of many LBS applications are only needed to be satisfied by different levels of accuracy \cite{LBSreport}.
This allows a user to misreport a different location $x_i'$ with $x_i'\neq x_i$ satisfying $D(x_i,x_i') \leq Q_i$ for privacy concerns,  bringing in degraded but acceptable service quality.
Such location perturbation or obfuscation is one of the common practices in location privacy protection mechanisms \cite{andres2013geo}.
To meet users' different service flexibilities, popular LBS apps such as Yelp also provides choices of different search ranges (0.5km, 2km, 5km, etc.) from a user's entered location to search for nearby PoIs \cite{Yelp}.

%
Provided with the crowdsourced cache, each user has access to all former users' searched PoI information. Then user $i$ benefits from the multi-user cooperation as long as he finds any former user $j$'s query $x_j'$ within distance $Q_i$, i.e., $\exists j=1,\cdots,i-1$ such that 
\begin{equation}
D(x_i,x_j') \leq Q_i.
\end{equation}

Fig.~\ref{Fig:Two users' cooperative reportings} illustrates our crowdsourced query scheme works in the following four steps. 
\begin{itemize}
\item 
Step 1: User $i \in \mathcal{U}$ first examines the data cached by former users for useful PoI information (if any) within distance $Q_i$ from his real location $x_i$.

\item
Step 2: User $i$ may continue to query the LBS platform with a location $x_i'$, which may be different from $x_i$.
Unlike the literature \cite{zhang2019caching,shokri_hiding_2014,peng2017collaborative,hu2018proactive,jung2017collaborative,cui2020cache}, we allow the user to still query even if he finds the shared PoIs useful. Actually, hiding without any query $x_i'$ is a special case of our query strategy here. 
The user may generally use different query strategies, depending on whether he finds cached PoIs useful or not. 

\item
Step 3: The platform returns PoI information $I(x_i')$ to the user upon request.

\item
Step 4:
User $i$ stores the query location $x_i'$ and the corresponding PoI information $I(x_i')$ in the cache to benefit latter users.
\end{itemize}


After observing $i-1$ former users' shared PoI information in the crowdsourced cache and querying the LBS platform at $x_i'$, user $i$ is believed to receive an expected privacy gain $\pi_i$. 
%
%
Next we will introduce any user $i$'s randomized strategy against the optimal inference attack to model the expected privacy gain $\pi_i$ at the $i$-th order.


\subsection{Users' Randomized Strategies against the Adversary}
Given former users’ query summary $\bm{x'}_{i-1}=(x_1',\cdots,x_{i-1}')$, it is user $i$'s turn to decide how to query. Depending on whether $\bm{x'}_{i-1}$ includes useful PoI information within walking distance $Q_i$ from $x_i$, user $i$ generally take two different query strategies in Step 2 above.
Mathematically, we define 
\begin{equation}
\label{Equ:X i IN:general}
\mathcal{X}_i^{in}(\bm{x'}_{i-1},Q_i)=
\cup_{j=1}^{i-1}
\{x \in \mathcal{M} |
D(x,x_j') \leq Q_i
\},
\end{equation} 
which summarizes the covered location set from earlier queries for user $i$'s real location $x_i$. 
If $x_i\in \mathcal{X}_i^{in}(\bm{x'}_{i-1},Q_i)$, user $i$ already finds the cached PoI information useful and meet his service constraint. 
Similarly, we define 
\begin{equation*}
\begin{aligned}
\mathcal{X}_i^{out}(\bm{x'}_{i-1},Q_i)
=
\mathcal{M}
\setminus
\mathcal{X}_i^{in}(\bm{x'}_{i-1},Q_i)
\end{aligned}
\end{equation*}
as the uncovered PoI location set for user $i$. 
%
If $x_i \in \mathcal{X}_i^{out}(\bm{x'}_{i-1},Q_i)$, user $i$ has to query with a new $x_i'$ nearby to meet his service constraint $D(x_i,x_i') \leq Q_i$.
Note that the first-query user 1 finds the cache empty with $\mathcal{X}_1^{in}=\emptyset$, $\mathcal{X}_1^{out}=\mathcal{M}$.

To determine the query location $x_i'$ for any given $x_i$, we generally use a conditional probabilistic distribution $f_i(x_i'|x_i)$ to denote user $i$'s randomized query strategy. It is a mapping probability from $x_i\in \mathcal{M}$ to $x_i'\in \mathcal{M}$, depending on $x_i\in \mathcal{X}_i^{in}(\bm{x'}_{i-1},Q_i)$ or $x_i\in \mathcal{X}_i^{out}(\bm{x'}_{i-1},Q_i)$.
It should be noted that in general no user will use a one-to-one deterministic query strategy as the adversary can easily infer his actual location $x_i$ from reported $x_i'$. 
\begin{defn}[User $i$'s randomized query strategy]
\label{Definition:f_i}
Depending on whether the cached PoIs are useful, we generally define user $i$'s query strategy to the LBS platform as
\begin{equation}
\label{Equ: f definition}
\begin{aligned}
&f_i(x_i'|x_i,\bm{x'}_{i-1},Q_i) \\=
&\begin{cases}
f_i^{in}(x_i'|x_i,\bm{x'}_{i-1},
Q_i
),  &\text{if } x_i \in \mathcal{X}_i^{in}(\bm{x'}_{i-1},Q_i), \\
f_i^{out}(x_i'|x_i,\bm{x'}_{i-1},Q_i), &\text{if } x_i \in \mathcal{X}_i^{out}(\bm{x'}_{i-1},Q_i).
\end{cases}
\end{aligned}
\end{equation} 
after observing $I(\bm{x'}_{i-1})$ shared by all $i-1$ former users. 
\end{defn}


As a special case of Definition \ref{Definition:f_i}, if the user hides from the LBS platform without any query as in \cite{zhang2019caching,shokri_hiding_2014,peng2017collaborative,hu2018proactive,jung2017collaborative,cui2020cache}, $f_i^{in}(x_i'|x_i,\bm{x'}_{i-1},Q_i)=\emptyset$ yet the adversary still learns $x_i \in \mathcal{X}_i^{in}(\bm{x'}_{i-1},Q_i)$.

Provided with the randomized query strategy in Definition \ref{Definition:f_i}, user $i$'s expected privacy gain also depends on the adversary's optimal inference attack. The adversary can access all former users' queries $\bm{x'}_{i-1}$ as well as user $i$'s query $x_i'$. Let $\hat{x}_i$ denote the adversary's optimal inference, which is a function of both $x_i'$ and $\bm{x'}_{i-1}$. 
To minimize the inference error, i.e., the expectation of random distance $D(\hat{x}_i,x_i)$ from the inferred location $\hat{x}_i$ to user $i$'s real location $x_i$, the adversary's optimal inference problem for user $i$ is given as
\begin{equation}
\label{Equ:inference error for user 2}
\min_{\hat{x}_i\in \mathcal{M}}
\int_{x_i\in \mathcal{M}} \operatorname{Pr}(x_i|x_i',\bm{x'}_{i-1})D(\hat{x}_i,x_i)dx_i,
\end{equation}
where $\operatorname{Pr}(x_i|x_i',\bm{x'}_{i-1})$ is a posterior probability of user $i$'s real location $x_i$. We will analyze this attack in detail later in Section \ref{Subsec:ad's opt infer}.

Let $\psi_i(x_i)$ denote the probability density function (PDF) of user $i$'s random location $x_i$, which is also known to the adversary by checking historical query data. Then user $i$'s expected privacy gain $\pi_i$ is defined as the expectation of random distance $D(x_i,\hat{x}_i(x_i'|\bm{x'}_{i-1}))$ from his real location $x_i$ to the adversary's optimal inference $\hat{x}_i$, by averaging over all possible $x_i$ and $x_i'$ by using two query strategies in Definition \ref{Definition:f_i}. That is,  
\begin{equation}
\label{Equ:user 2's expected privacy-latter users}
\begin{aligned}
\pi_i=\int_{x_i' \in \mathcal{M}}\int_{x_i \in \mathcal{X}_i^{in}(\bm{x'}_{i-1},Q_i)} &\psi_i(x_i) f_i^{in}(x_i'|x_i,\bm{x'}_{i-1},Q_i) \\ &D(x_i,\hat{x}_i(x_i'|\bm{x'}_{i-1}))dx_i dx_i' \\
+\int_{x_i' \in \mathcal{M}}\int_{x_i \in \mathcal{X}_i^{out}(\bm{x'}_{i-1},Q_i)} &\psi_i(x_i) f_i^{out}(x_i'|x_i,\bm{x'}_{i-1},Q_i) \\ &D(x_i,\hat{x}_i(x_i'|\bm{x'}_{i-1})) dx_i dx_i'  \\
=\int_{x_i'\in \mathcal{M}}\int_{x_i\in \mathcal{M}}&\operatorname{Pr}(x_i,x_i'|\bm{x'}_{i-1}) \\
&D(x_i,\hat{x}_i(x_i'|\bm{x'}_{i-1})) dx_i dx_i'.
\end{aligned}
\end{equation}
The last equality is due to the product rule on conditional probability: $\operatorname{Pr}(x_i,x_i')=\psi_i(x_i) f_i(x_i'|x_i)$.
Next we are ready to model the strategic interaction between $N$ cooperative users and the adversary as an adversarial Bayesian game. 

%

\subsection{Adversarial Bayesian Game Formulation}
Based on the adversary's optimal inference attack formulation in (\ref{Equ:inference error for user 2}), we formally model the strategic interaction between any user $i$ and the adversary as an adversarial Bayesian game with two stages:
\begin{itemize}
\item 
In Stage I, user $i$ decides a probabilistic strategy $\bm{f}_i(x_i'|x_i,\bm{x'}_{i-1})=\{f_i^{in}(x_i'|x_i,\bm{x'}_{i-1},Q_i),f_i^{out}(x_i'|x_i,\bm{x'}_{i-1},Q_i)\}$ in (\ref{Equ: f definition}) to query the LBS platform. The objective is to maximize his expected privacy gain in (\ref{Equ:user 2's expected privacy-latter users}), depending on whether he finds the cached PoI information useful or not. 

\item
In Stage II, without the knowledge of user $i$'s real location $x_i$ but prior distribution $\psi_i(x_i)$, the adversary launches its optimal Bayesian inference attack to infer user $i$’s location as $\hat{x}_i$.
To provide robust privacy preservation (\cite{wu2020zero,chattopadhyay2021robustness}), we look at the challenging case that the adversary has access to the cached data $\bm{x}_{i-1}'$ upon user $i$'s query, and knows users' privacy gain function in (\ref{Equ:user 2's expected privacy-latter users}). 
After observing user $i$'s query location $x_i'$ as well as former users' queries $\bm{x'}_{i-1}$, its objective is to minimize its inference error in  (\ref{Equ:inference error for user 2}).

\end{itemize}

In our multi-user crowdsourced query scheme, there is no conflict between any two users, as each user sequentially demands PoI and maximizes the individual privacy gain.

So far we have modelled the strategic interaction between cooperative users and the 
adversary as a two-stage Bayesian game, where users move first by making the query against the adversary's inference attack. Next in Section \ref{Sec: pri coop against ad inference}, we will analyze the Bayesian game by backward induction.

\section{Multi-user Cooperation against Adversarial Inference Attack}
\label{Sec: pri coop against ad inference}
In this section, we first analyze the adversary's optimal inference attack $\hat{x}_i$. Then we proactively design each user $i$'s two different query strategies $f_i^{in}$ and  $f_i^{out}$ in (\ref{Equ: f definition}), by taking the adversary's optimal attack response $\hat{x}_i$ into account. 
%
%
Finally, we discuss on whether to hide from the LBS platform or not if the user already finds useful information in the cache.

\subsection{Adversary's Optimal Bayesian Inference Attack}
\label{Subsec:ad's opt infer}

To ensure reliable performance for the user, we consider the worst-case of the fully informed adversary, which is a standard approach of modelling  robust privacy defence \cite{chattopadhyay2021robustness,wu2020zero}.
By accessing the cache, 
any cooperative peers and/or the historical data from the LBS platform, the adversary can easily have full knowledge of the reported data $\bm{x}_{i}'$, each user’s privacy gain function as in (\ref{Equ:user 2's expected privacy-latter users}),
all users' privacy objective functions 
and their location distributions $\psi_i(x_i)$ ($i=1,\cdots,N$). 
Thus it can estimate users' query strategies  $f_i(x_i'|x_i,\bm{x}_{i-1}')$ on their behalves, and launch the optimal Bayesian inference attack
 of each user’s real location.

After observing user $i$'s query location $x_i'$ as well as $i-1$ former users' queries $\bm{x'}_{i-1}$, the adversary updates the posterior probability of user $i$'s real location $x_i$ below: 
\begin{equation}
\label{Equ:posterior distribution}
\begin{aligned}
\operatorname{Pr}(x_i|x_i',\bm{x'}_{i-1})
&=\frac{\operatorname{Pr}(x_i,x_i'|\bm{x'}_{i-1})}{\operatorname{Pr}(x_i'|\bm{x'}_{i-1})} \\
&=
\frac{\psi_i(x_i)f_i(x_i'|x_i,\bm{x'}_{i-1},Q_i) }
{
\int_{x_i \in \mathcal{M}}
\psi_i(x_i)f_i(x_i'|x_i,\bm{x'}_{i-1},Q_i) dx_i
},
\end{aligned}
\end{equation}
where the denominator is the integral over the whole location region $\mathcal{M}$. The strategy $f_i$ equals $f_i^{in}(x_i'|x_i,\bm{x'}_{i-1},Q_i)$ for $x_i \in \mathcal{X}_i^{in}(\bm{x'}_{i-1},Q_i)$ or $f_i^{out}(x_i'|x_i,\bm{x'}_{i-1},Q_i)$ for $x_i \in \mathcal{X}_i^{out}(\bm{x'}_{i-1},Q_i)$ in (\ref{Equ: f definition}).
As a special case, for the first user to arrive and query with $\mathcal{X}_1^{in}=\emptyset$ and $\mathcal{X}_1^{out}=\mathcal{M}$, (\ref{Equ:posterior distribution}) reduces to 
\begin{equation*}
	\operatorname{Pr}(x_1|x_1')=\frac{\operatorname{Pr}(x_1,x_1')}{\operatorname{Pr}(x_1')}
	=
	\frac{\psi_1(x_1)f_1^{out}(x_1'|x_1) }{\int_{x_1 \in \mathcal{M}}\psi_1(x_1)f_1^{out}(x_1'|x_1) dx_1}.
\end{equation*}

By substituting the posterior probability in (\ref{Equ:posterior distribution}) to (\ref{Equ:inference error for user 2}),  we can analyze the adversary's optimal inference problem for user $i$ in
(\ref{Equ:inference error for user 2}).
By taking the adversary's inference into consideration, we next reformulate each user's obfuscated query problem in (\ref{Equ:user 2's expected privacy-latter users}) as a max-min optimization problem.

\subsection{Max-min Problem Formulation and Simplification}

By taking the adversary’s inference problem in (\ref{Equ:inference error for user 2}) using (\ref{Equ:posterior distribution}) into consideration, we rewrite any user $i$'s expected privacy gain $\pi_i$ from (\ref{Equ:user 2's expected privacy-latter users}) to:
\begin{multline}
\label{Equ: user i's expected privacy-average over x_i'}
\pi_i=
\int_{x_i' \in \mathcal{M}}\operatorname{Pr} (x_i'|\bm{x'}_{i-1}) \\
\min_{\hat{x}_i \in \mathcal{M}} 
\int_{x_i \in \mathcal{M}} \operatorname{Pr}(x_i|x_i',\bm{x'}_{i-1})D(\hat{x}_i,x_i) dx_i
dx_i' \\
=
\int_{x_i' \in \mathcal{M}}
\min_{\hat{x}_i \in \mathcal{M}} 
\int_{x_i \in \mathcal{M}} \psi_i(x_i) f_i(x_i'|x_i,\bm{x'}_{i-1},Q_i) \\
D(\hat{x}_i,x_i) dx_i   dx_i',
\end{multline}
which integrates over any possible $x_i'$.
Both equalities in (\ref{Equ: user i's expected privacy-average over x_i'}) hold due to the product rule on conditional probability: $ \operatorname{Pr}(x_i,x_i')=\operatorname{Pr} (x_i')\operatorname{Pr}(x_i|x_i')$ for the first equality and $\operatorname{Pr}(x_i,x_i')=\psi_i(x_i) f_i(x_i'|x_i)$ for the second equality.

User $i$ aims to optimize its expected privacy gain $\pi_i$ in (\ref{Equ: user i's expected privacy-average over x_i'}), by considering his PoI location requirement within distance $Q_i$ from $x_i$: 
\begin{equation}
\label{Opt:user 2's reporting prob-cont}
\begin{aligned}
\max 
\int_{x_i'\in \mathcal{M}}
\min_{\hat{x}_i\in \mathcal{M}} 
\int_{x_i\in \mathcal{M}} \psi_i(x_i) f_i(x_i'|x_i,\bm{x'}_{i-1},Q_i) \\ D(\hat{x}_i,x_i) dx_i   dx_i' \\
\text{s.t. }
f_i^{out}(x_i'|x_i,\bm{x'}_{i-1},Q_i)=0, 
\forall (x_i, x_i') \in  \\
\left\lbrace(x_i, x_i')|  
x_i\in \mathcal{X}_i^{out}(\bm{x}_{i-1}',Q_i), 
x_i'\in \mathcal{M},
D(x_i',x_i)> Q_i \right\rbrace,\\ 
\int_{x_i'\in \mathcal{M}}f_i(x_i'|x_i,\bm{x'}_{i-1},Q_i)dx_i'=1, \forall x_i \in \mathcal{M}, \\
var: f_i(x_i'|x_i,\bm{x'}_{i-1},Q_i) \text{\emph{ in }} (\ref{Equ: f definition}), \forall x_i, x_i' \in \mathcal{M}.
\end{aligned}
\end{equation}
The first constraint of (\ref{Opt:user 2's reporting prob-cont}) is to meet the service requirement of user $i$. If he finds his real location $x_i\in \mathcal{X}_i^{in}(\bm{x'}_{i-1},Q_i)$, this requirement is met regardless of his query strategy $f_i^{in}$. Otherwise, if $x_i\in \mathcal{X}_i^{out}(\bm{x'}_{i-1},Q_i)$, we should ensure that user $i$ will not report a location $x_i'$ with more than distance $Q_i$ away from $x_i$.
Thus, we require $D(x_i' , x_i)\leq Q_i$, or equivalently zero probability $f_i^{out}(x_i'|x_i,\bm{x'}_{i-1},Q_i)=0$ for $D(x_i' , x_i)> Q_i$.

As the 
location region $\mathcal{M}$ is continuous, it is difficult to solve two continuous variable functions $f_i^{in}(x_i'|x_i,\bm{x'}_{i-1},Q_i)$ and $f_i^{out}(x_i'|x_i,\bm{x'}_{i-1},Q_i)$ in (\ref{Opt:user 2's reporting prob-cont}). 
Instead, we propose the following alternative. 
\begin{prop}
\label{Prop:prob reformulation-discrete}
By equally partitioning the 
continuous location set $\mathcal{M}$ into $M$ discrete grids, we simplify the objective of Problem (\ref{Opt:user 2's reporting prob-cont}) as: 
\begin{multline}
\label{Opt:user 2's reporting prob-discrete}
\max
\sum_{x_i'}
\min_{\hat{x}_i} 
\sum_{x_i \in \mathcal{M}} \psi_i(x_i) f_i(x_i'|x_i,\bm{x'}_{i-1},Q_i) D(\hat{x}_i,x_i), 
\end{multline} 
then the problem (\ref{Opt:user 2's reporting prob-cont}) for each user $i$ becomes a linear program (LP) with computational complexity $\mathcal{O}(M^7)$.
\end{prop}

The time complexity of (\ref{Opt:user 2's reporting prob-discrete}) is derived as inspired by \cite{karmarkar1984new}, which solved an LP problem with $n$ variables and $m$ constraints by an interior algorithm within complexity $\mathcal{O}(m^{3/2}n^2)$.
By using some toolboxes (e.g., Optimization Toolbox in MATLAB), we can solve our LP problem numerically and our solution approaches the optimum as $M$ goes to infinity to reduce the discretization error. 
Despite of the error, this LP may not be solvable for a large-scale spatial region and we will propose an approximate obfuscated query solution on the original problem (\ref{Opt:user 2's reporting prob-cont}) later in Section \ref{Sec:approximation1D}.

\subsection{To hide or not when $x_i\in \mathcal{X}_i^{in}(\bm{x'}_{i-1},Q_i)$}
\label{Subsec: hiding compared with opt}

Recall that in the existing multi-user privacy preservation schemes in \cite{zhang2019caching,shokri_hiding_2014,peng2017collaborative,hu2018proactive,jung2017collaborative,cui2020cache}, users query the LBS platform only if the target PoI information is not found in the cache. Their hiding-from-the-LBS idea leads to
$f_i^{in}=\emptyset$ in (\ref{Equ: f definition}) in our problem formulation as there is no query for $x_i\in \mathcal{X}^{in}(\bm{x'}_{i-1})$. In this case, the adversary can still infer user $i$'s location in the covered region $\mathcal{X}^{in}(\bm{x'}_{i-1})$.
We can similarly formulate a $\max$-$\min$ problem for the hiding scheme as  (\ref{Opt:user 2's reporting prob-cont}), by only limiting the decision variable function to $f_i^{out}$ for the case of $x_i \in \mathcal{X}_i^{out}(\bm{x'}_{i-1},Q_i)$.
Let $\mathbb{E} \pi^{opt}=\frac{1}{N} \sum_i \pi_i^{opt}$ denote {\textbf{the average user's privacy gain under the optimal solution}} to problem (\ref{Opt:user 2's reporting prob-cont}). 
Let $\mathbb{E} \tilde{\pi}^{opt}$ denote the counterpart from problem (\ref{Opt:user 2's reporting prob-discrete}) after discretization. Then when $M$ is large enough, $\mathbb{E} \tilde{\pi}^{opt}$ approaches $\mathbb{E} \pi^{opt}$.
We also let $\mathbb{E} \pi^{hide}$ denote {\textbf{the average user's privacy gain under the existing hiding schemes}} as in \cite{zhang2019caching,shokri_hiding_2014,peng2017collaborative,hu2018proactive,jung2017collaborative,cui2020cache}. 
Next we analytically compare the existing hiding solution with to our always-query solution.

\begin{figure}[!t]
	\centering
	\includegraphics[width=2.4 in]{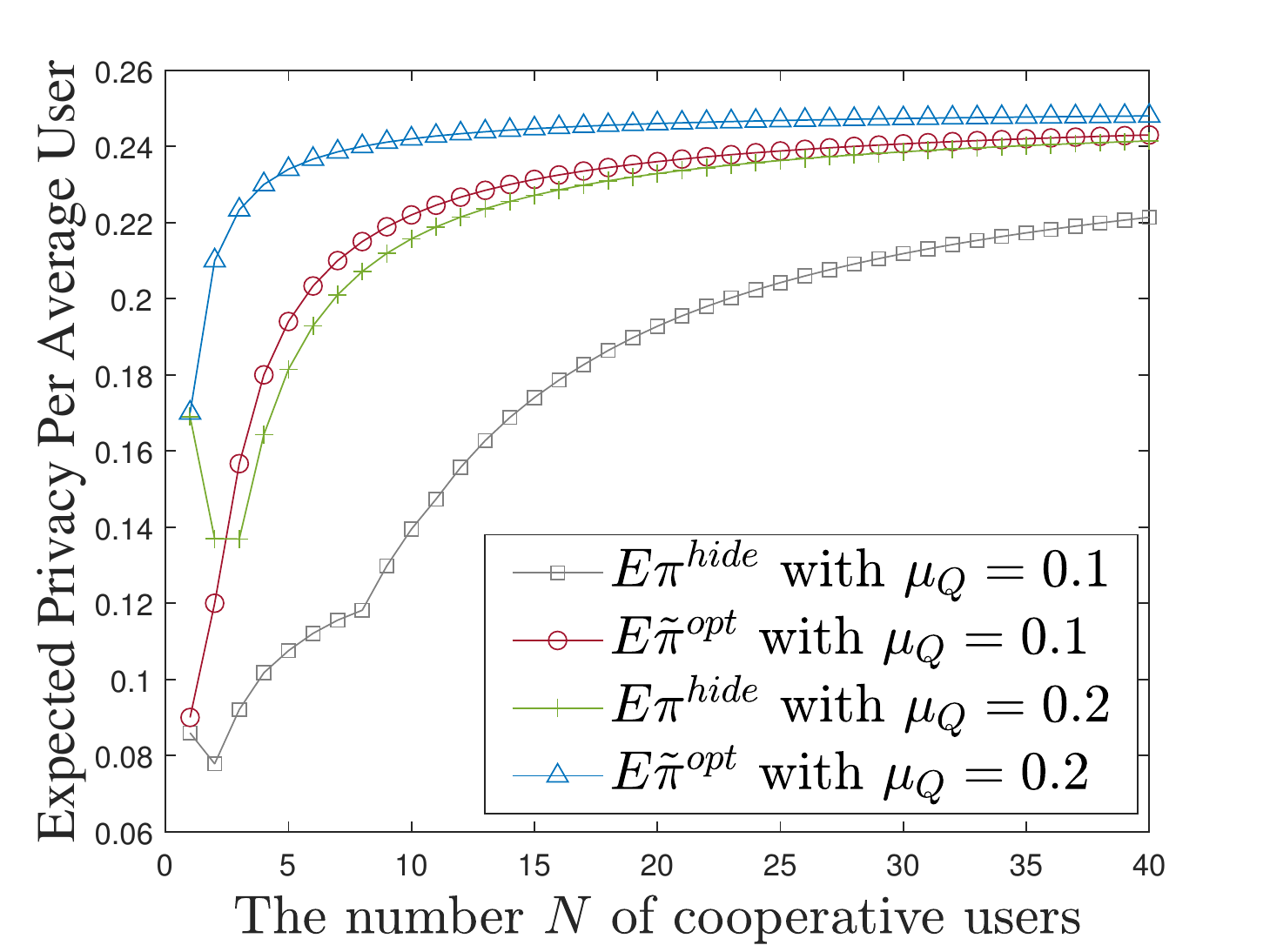}
\caption{Performance comparison between our strategy's performance $\mathbb{E}\tilde{\pi}^{opt}$ to Problem (\ref{Opt:user 2's reporting prob-discrete}) and the existing hiding strategy's performance $\mathbb{E} \pi^{hide}$ in the literature (\cite{zhang2019caching,shokri_hiding_2014,peng2017collaborative,hu2018proactive,jung2017collaborative,cui2020cache}). We set each user $i$'s service flexibility constraint $Q_i$ to follow the i.i.d. truncated normal distribution with mean $\mu_Q=0.1$ or $0.2$ under the standard deviation $\sigma=0.1$ among $N$ users.The discretization size for $\mathcal{M}$ is $M=20$.
\label{Fig:hide performance}}
\end{figure}

\begin{lemma}
\label{lemma:hide}
$\mathbb{E} \pi^{opt} > \mathbb{E} \pi^{hide}$ always holds for $N>1$ and it is always beneficial for each user to strategically query the LBS platform with $f_i^{in}\neq \emptyset$ in (\ref{Equ: f definition}). 
\end{lemma}

Lemma \ref{lemma:hide} answered our first key question in the introduction that even if the user already gets help from the crowdsourced cache, querying the LBS platform strategically helps improve the privacy gain further. 
The intuitions behind are explained in the two following aspects: 
\begin{itemize}
\item 
\emph{Adding confusion to the adversary:} When the user always queries, the adversary does not know $x_i \in \mathcal{X}_i^{in}(\bm{x'}_{i-1},Q_i)$
or $x_i \in \mathcal{X}_i^{out}(\bm{x'}_{i-1},Q_i)$.
The two randomized strategies $f_i^{in}(x_i'|x_i,\bm{x'}_{i-1},Q_i)$ and $f_i^{out}(x_i'|x_i,\bm{x'}_{i-1},Q_i)$ in (\ref{Equ: f definition}) jointly confuse the adversary. 
\item 
\emph{More PoIs to benefit latter users:} Our always-query strategy returns more PoI data in the cache for latter users to take advantage of. Thus the covered PoI location sets are enlarged for latter users to use.
\end{itemize}

Besides the analytical comparison in Lemma \ref{lemma:hide}, we run simulations in Fig.~\ref{Fig:hide performance} to empirically compare the tractable performance $\mathbb{E}\tilde{\pi}^{opt}$ by solving Problem (\ref{Opt:user 2's reporting prob-discrete}) with $\mathbb{E} \pi^{hide}$ for any user number $N$.
Here all users' service flexibility constraints $Q_i$ follow the i.i.d. truncated normal distribution with the minimum value $Q_i \geq 0$. 
Given the same standard deviation, we examine the performances under different mean values $\mu_Q$ of $Q_i$ distribution. 

Due to the PoI sharing benefit in multi-user cooperation, all privacy performance curves (no matter for hiding strategy or not) are generally increasing in user number $N$. The expected privacy gain increases in the mean value $\mu_Q$ of $Q_i$, as a greater flexibility helps users add obfuscation to their queries strategically for a better privacy gain. 
An average user's expected privacy gain $\mathbb{E}\tilde{\pi}^{opt}$ obviously improves under our always-query strategy, as compared to the traditional $\mathbb{E} \pi^{hide}$ with hiding. The performance advantage becomes obvious for a non-small $N$, as the our always-query strategy creates the maximum obfuscation to the adversary and creates more PoIs to share among users. 

On the other hand, the hiding strategy from the LBS platform may expose the user's real location to the cache-covered location set $\mathcal{X}_i^{in}$ and facilitate the adversary's inference attack.
Note that in Fig.~\ref{Fig:hide performance} $\mathbb{E} \pi^{hide}$ even decreases from $N=1$ to $2$ for both curves $\mu_Q=0.1$ and $\mu_Q=0.2$. 
It is because that the second user's possible hiding exposes the user's narrowed location region in $\mathcal{X}_2^{in}(x'_1)$ and helps the adversary to locate his infer $\hat{x}_2$ around the former user 1's query $x_1'$.
Only after more users' joining can the cooperation benefits outweight the hiding disadvantage.

\section{Approximate Obfuscated Query Scheme}
\label{Sec:approximation1D}
To make the problem (\ref{Opt:user 2's reporting prob-cont}) solvable for a large-scale continuous region and provide clean engineering insights, this section presents our approximate obfuscated query scheme in closed-form. Without much loss of generality, we first assume the users' location set $\mathcal{M}$ to be a normalized 1D line interval $\mathcal{M}=[0,1]$. We will similarly extend our analysis and solution to a 2D plane in Section \ref{Subsec:extension 2D} later. We also assume that each user is equally located at any point in interval $\mathcal{M}$ by following an i.i.d. uniform distribution with PDF $\psi_i(x_i)=1$. Similar results and algorithms can be extended to arbitrary location distributions.

Next we will start with analyzing the first user's approximate query strategy facing an empty cache in Section \ref{Subsec:first user}. Then we will analyze for the second user and latter users in Sections \ref{Subsec:user 2 coop 1D appr} and \ref{Subsec:latter users}, respectively.
Instead of deciding continuous functions $f_i^{in}$ and $f_i^{out}$ in (\ref{Equ: f definition}) for user $i \in \mathcal{N}$ over the location set $\mathcal{M}$, our key approximation idea is to reduce each of them to be the randomization of only two location points in two opposite spatial directions from $x_i$.

\subsection{The First User's Random Query Strategy}
\label{Subsec:first user}

\begin{figure}[!t]
\centering
\includegraphics[width=3 in]{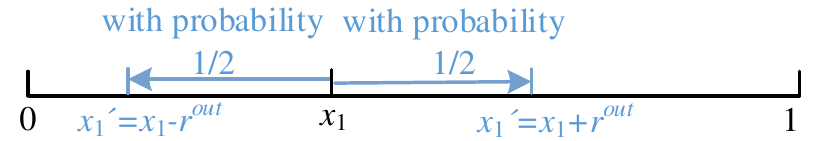}
\caption{User 1's query strategy $f_1^{out}(x_1'|x_1,\emptyset)$ with two randomized queries  $x_1'=|x_1-r^{out}|$ or $x_1'=1-|x_1-1+r^{out}|$ with equal probability $\frac{1}{2}$, in two spatial directions of $x_1$.}
\label{Fig:1D user 1's reporting}
\end{figure}

First, we discuss the approximate query strategy for the first user 1 to demand PoIs, who observes an empty cache (i.e., $ \mathcal{X}_1^{in}=\emptyset$) and its strategy only includes $f_1^{out}$ according to Definition \ref{Definition:f_i}. To simplify the continuous query strategy $f_1^{out}(x_1'|x_1,\emptyset)$, we approximate user 1's strategy to a two-sided random query scheme as:
\begin{align}
\label{Equ: f1 out}
f_1^{out}(x_1'|x_1,\emptyset) = 
\begin{cases} 
\frac{1}{2}, &\text{if } x_1'=|x_1-r^{out}|,\\
\frac{1}{2}, &\text{if } x_1'=1-|x_1-1+r^{out}|, 
\end{cases} 
\end{align}
where the absolute term appears to keep query location $x_1'$ within the interval $\mathcal{M}$. Such binary approximation is the simplest but fundamental
way to replace the complicated randomization function $f_1^{out}$.
To maximizing $\pi_1$ in (\ref{Opt:user 2's reporting prob-cont}), user 1's decision now changes from $f_1^{out}(x_1'|x_1,\emptyset)$ to $r^{out}$ only.

\begin{prop} \label{Prop: user 1 strategy}
The {\textbf{optimal obfuscation distance}} in (\ref{Equ: f1 out}) to maximize the first user’s expected privacy gain is
\begin{equation}
	\label{Equ:r out}
	r^{out}=\min(Q_1,\frac{1}{2}).
\end{equation} 
His maximum expected privacy gain is $\pi_1^{appr}=\min(Q_1-{Q_1^2},\frac{1}{4})$.
\end{prop}

To add maximum obfuscation, user 1 reports the farthest possible query point $x_1'$ with maximum distance $Q_1$ from $x_1$, while merely meeting the service constraint. Even knowing $r^{out}$, the adversary is unsure about $x_1$ given the binary randomization.

\subsection{The Second User's Random Query Strategy}
\label{Subsec:user 2 coop 1D appr}

After observing the historical query data $\bm{x}_{i-1}'$ and his covered PoI location set   
$$\mathcal{X}_i^{in}(\bm{x}_{i-1}')=(\cup_{j=1}^{i-1} [x_j'-Q_i,x_j'+Q_i]) \cap [0,1],$$
user $i\in\{2,...,N\}$ needs to decide $f_i^{in}(x_i'|x_i,\bm{x'}_{i-1},Q_i)$ and $f_i^{out}(x_i'|x_i,\bm{x'}_{i-1},Q_i)$ in Definition \ref{Definition:f_i}. Similar to (\ref{Equ: f1 out}), we apply the following {\textbf{binary approximation}} of users' obfuscated query strategy.  

\begin{defn}[Binary approximation for obfuscated query]
\label{Definition:binary reporting}
If a user $i$ finds the cached PoI information useful (i.e., $x_i\in \mathcal{X}_i^{in}(\bm{x'}_{i-1},Q_i)$), we approximate his query strategy to randomization of two symmetric points on the two sides of his real location $x_i$: 
	\begin{align}
		\label{Equ: f_in def}
		f_i^{in}(x_i'|x_i,\bm{x'}_{i-1},Q_i) = 
		\begin{cases} 
			\frac{1}{2}, &\text{if } x_i'=|x_i-r_i^{in}|,\\
			\frac{1}{2}, &\text{if } x_i'=1-|x_i-1+r_i^{in}|.
		\end{cases} 
	\end{align}
	Otherwise, if $x_i\in \mathcal{X}_i^{out}(\bm{x'}_{i-1},Q_i)$, 
	\begin{align}
		\label{Equ: f_out def}
		f_i^{out}(x_i'|x_i,\bm{x'}_{i-1},Q_i) = 
		\begin{cases} 
			\frac{1}{2}, &\text{if } x_i'=|x_i-r^{out}(Q_i)|,\\
			\frac{1}{2}, &\text{if } x_i'=1-|x_i-1+r^{out}(Q_i)|.
		\end{cases} 
	\end{align}
\end{defn}

Note that this definition also holds for user 1, as he always finds $x_1\in \mathcal{X}_1^{out}(\bm{x'}_{0}=\emptyset)=\mathcal{M}$.
Such binary approximation is the simplest but fundamental way to replace the complicated randomization functions $f_i^{in}$ and $f_i^{out}$.
For any user $i$ demanding PoIs later than user 1, (e.g., user 2), if $x_i\in \mathcal{X}_i^{in}(\bm{x'}_{i-1},Q_i)$, he should take different strategies of $r_i^{in}$ in (\ref{Equ: f_in def}) according to prior queries $\bm{x}_{i-1}'$. If $x_i\in \mathcal{X}_i^{out}(\bm{x'}_{i-1},Q_i)$, as in user 1's case, to add maximum obfuscation, we suppose user $i$ reports the farthest possible query point $x_i'$ with maximum distance $Q_i$ from $x_i$. 

For ease of exposition, we separate the analysis for user 2 and the following users and in the rest of this subsection we focus on user 2 who is the second to query to explain. 
Next we optimize $r_2^{in}$ for maximizing user 2's expected privacy gain $\pi_2$.
To avoid the trivial case that user 2 has sufficient flexibility or large $Q_2$ in the first constraint of (\ref{Opt:user 2's reporting prob-cont}) 
to arbitrarily misreport his location without service loss, we consider a challenging case with $Q_2<1/11$ here.  
The analysis can be extended for a larger $Q_2$  with a better privacy performance. 

\begin{prop} 
\label{Prop:r_{in}}
Assuming $Q_2<1/11$, even if user 2 finds the PoI information shared by user 1 useful (i.e., $D(x_1',x_2)\leq Q_2$), it is optimal for him to misreport $x_2'$ with distance $r_2^{in}(x_1')$ away from $x_2$ in Definition \ref{Definition:binary reporting} when querying  the LBS platform, where the obfuscation distance is given by 
\begin{equation}
\label{Equ:r_in}
r_2^{in}(x_1')=
\begin{cases}
1-Q_2, &\text{ if }0 \leq x_1' \leq Q_2, \\
1-x_1', &\text{ if }Q_2 < x_1' \leq \frac{1}{2}, \\
x_1', &\text{ if }\frac{1}{2} < x_1' \leq 1-Q_2, \\
1-Q_2, &\text{ if }1-Q_2 < x_1' \leq 1.
\end{cases}
\end{equation}
\end{prop}

As our binary approximation solution is in closed-form, we manage to reduce the computational complexity of deciding each user's query strategy from $\mathcal{O}(M^7)$ in Problem (\ref{Opt:user 2's reporting prob-discrete}) to $\mathcal{O}(1)$ here.
Surprisingly, notice that $r_2^{in}$ in (\ref{Equ:r_in}) is non-increasing in user 2's service flexibility $Q_2$.
This tells that the user with a  greater service flexibility should query the LBS platform with less obfuscated location, for strategically confusing the adversary.

\begin{figure}[!t]
	\centering
	\includegraphics[width=2.4 in]{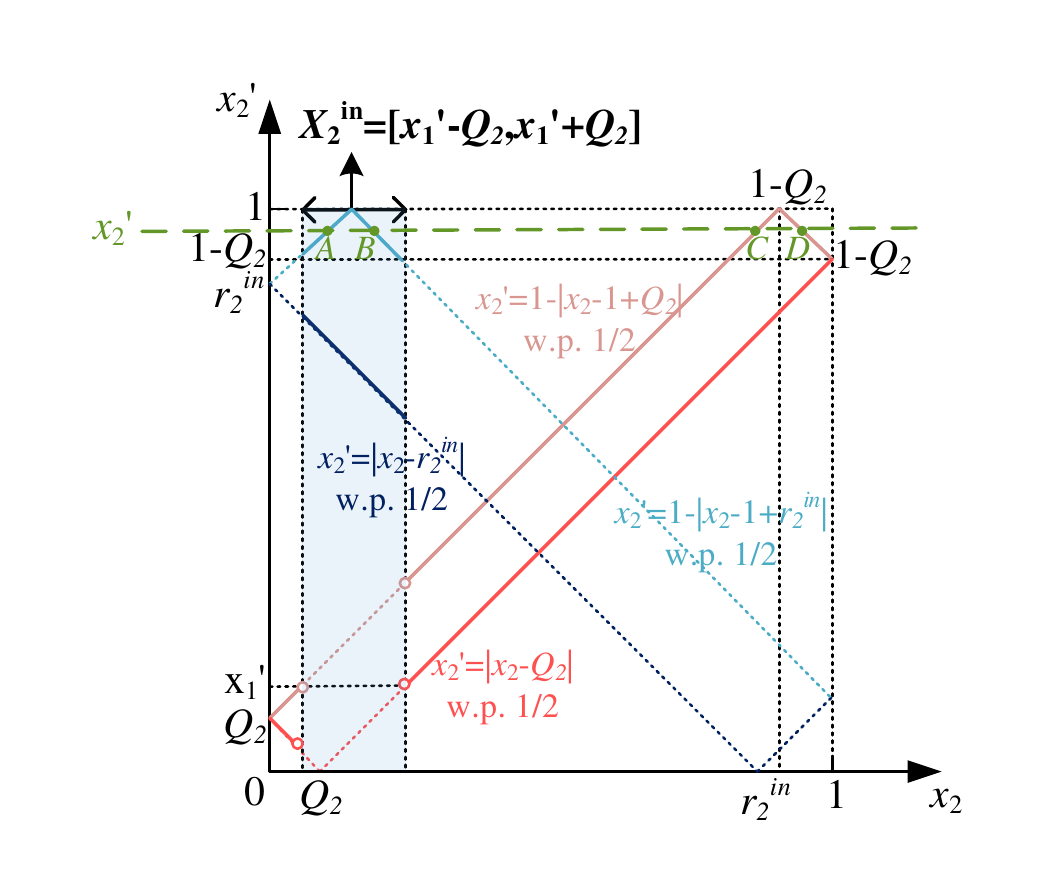}
	\caption{Example of $Q_2<x_1' \leq \frac{1}{2}$
		for explaining user 2's approximate query strategy: $f_2^{in}(x_2'|x_2,x_1')$ with $r_2^{in}$ in (\ref{Equ:r_in}) in two blue solid lines and $f_2^{in}(x_2'|x_2,x_1')$ in two red solid lines.}
	\label{Fig:2User-example}
\end{figure}

To better explain the implication of $r_2^{in}$ design in (\ref{Equ:r_in}), we present Fig.~\ref{Fig:2User-example} to give an example of user 2's query strategy.
In this example, user 1's realized query satisfies $Q_2<x_1' \leq \frac{1}{2}$, and user 2 has a covered PoI interval $\mathcal{X}_2^{in}(x_1')=[x_1'-Q_2,x_2'+Q_2]$ thanks to user 1's query.
If $x_2 \in \mathcal{X}_2^{in}(x_1')$, user 2's service constraint is met and he will query on the two sides of $x_2$ randomly with $f_2^{in}(x_2'|x_2,x_1')$ in (\ref{Equ: f_in def}) with $r_2^{in}$ close to 1 in (\ref{Equ:r_in}). This strategy $f_2^{in}(x_2'|x_2,x_1')$ is shown in two blue solid lines (i.e., $x_2'=|x_2-r_2^{in}|$ and $x_2'=1-|x_2-1+r_2^{in}|$) which are with equal probability $\frac{1}{2}$. 
If $x_2 \in \mathcal{X}_2^{out}(x_1')$, user 2 cannot benefit from the cooperation and he will query with $f_2^{out}(x_2'|x_2,x_1')$ in (\ref{Equ: f_out def}) with $r^{out}(Q_2)=Q_2$ to satisfy the service requirement.
This strategy $f_2^{out}(x_2'|x_2,x_1')$ is shown in two red solid lines (i.e., $x_2'=|x_2-Q_2|$ and $x_2'=1-|x_2-1+Q_2|$) which are with equal probability $\frac{1}{2}$.

In this example, when the adversary observes user 2's query $x_2'$ as highlighted by the green dash line in Fig.~\ref{Fig:2User-example}, it has four intersections $A,B,C,D$ with user 2's query strategy.
This leads to four possible real locations of inferring $x_2 $ with equal probabilities: the adversary is even not sure whether user 2 is in the covered PoI interval $\mathcal{X}_2^{in}$ (for points $A,B$) or in $\mathcal{X}_2^{out}$ (for points $C,D$).
Though applying binary approximation, our randomized query strategy makes it difficult for the attacker to launch the optimal inference attack in (\ref{Equ:inference error for user 2}).
It is also better than hiding strategy in the literature (\cite{zhang2019caching,shokri_hiding_2014,peng2017collaborative,hu2018proactive,jung2017collaborative,cui2020cache}), where the adversary immediately infers $x_2$ in the narrow interval $\mathcal{X}_2^{in}$ (with points $A$, $B$) without much obfuscation.

Given the latter user 2's query strategy is adjusted according to user 1's reports $x_1'$, the following Corollary \ref{Corol:Pi_2 appr reaches the min when x1'=d/2} tells the impact of $x_1'$ on user 2's expected privacy gain.
\begin{corol}
\label{Corol:Pi_2 appr reaches the min when x1'=d/2}
User 2's expected privacy gain $\pi_2^{appr}(x_1')$ in the approximate obfuscated query scheme in Definition \ref{Definition:binary reporting} using $r_2^{in}$ in (\ref{Equ:r_in}) reaches its minimum when user 1's query point is in the middle of location set $\mathcal{M}$ (i.e., $x_1'=\frac{1}{2}$).
\end{corol}
When $x_1'$ is at the centre of the location set $\mathcal{M}$, the four possible points for inferring $x_2$ (see points $A$, $B$, $C$ and $D$ in Fig.~\ref{Fig:2User-example}) will be close and user 2 cannot add much obfuscation to its query. 
This leads to a smaller privacy gain with less confusion to the adversary.

Compared with the existing hiding schemes \cite{zhang2019caching,shokri_hiding_2014,peng2017collaborative,hu2018proactive,jung2017collaborative,cui2020cache}, we can also show the performance of our obfuscation scheme.
\begin{corol}
\label{Corol:hide for user 2}
$\pi_2^{appr} > \pi_2^{hide}$ always holds for the second user. 
\end{corol}
Given the adversary knows the user's full strategy, when observing no query from user 2, it can narrow the location region for user 2 to $\mathcal{X}_2^{in}$ under the hiding-from-the-LBS scheme. 

\subsection{Latter Users' Random Query Strategy}
\label{Subsec:latter users}

For any arbitrary user $i\geq 3$, we similarly apply binary approximation in Definition \ref{Definition:binary reporting} to simplify the max-min problem (7) and derive his privacy gain $\pi_i^{appr}$, by optimizing $r_i^{in}$ in (\ref{Equ: f_in def}). 
Though more involved, we still follow the backward induction by first analyzing the adversary's optimal inference $\hat{x}_i(x_i'|\bm{x'}_{i-1})$ and then maximizing user $i$'s expected privacy gain $\pi_i^{appr}$.

\subsubsection{Adversary's Optimal Inference Attack under Definition \ref{Definition:binary reporting}}
Similar to Section \ref{Subsec:ad's opt infer}, the adversary can infer user $i$'s approximated query strategy in Definition \ref{Definition:binary reporting} including $r_i^{in}$ in (\ref{Equ:r_in}) and $r_i^{out}$ in (\ref{Equ:r out}). 
Given the adversary's posterior probability $\operatorname{Pr}(x_i|x_i',\bm{x}_{i-1}')$ of user $i$'s location $x_i$ in (\ref{Equ:posterior distribution}), Lemma \ref{Lemma:ad's possible real locations} below summarizes all possible real locations $\hat{\mathcal{X}}_i$ from the perspective of the adversary.

\begin{lemma}
\label{Lemma:ad's possible real locations}
After observing user $i$'s query $x_i'$ and former queries $\bm{x}_{i-1}'$ in the cache, the adversary believes the location $x_i$ of user $i\geq 3$ is equally likely to appear in the following set: 
\begin{multline}
\label{Equ: x_poss}
\hat{\mathcal{X}}_i(r_i^{in}, r^{out}, x_i', \bm{x}_{i-1}')\\
=
\{
x_i \in \mathcal{X}_i^{in}(\bm{x'}_{i-1},Q_i)|x_i'=|x_i-r_i^{in}|  
\text{ or } 1-|x_i-1+r_i^{in}|
\} \\
\cup
\{
x_i \in \mathcal{X}_i^{out}(\bm{x'}_{i-1},Q_i)|x_i'=|x_i-r_i^{out}|  
\text{ or } 1-|x_i-1+r_i^{out}|
\}.
\end{multline}
The adversary's optimal guess of $x_i$ is the mean of these possible locations: 
\begin{equation}
	\label{Equ: ad infer}
	\hat{x}_i(r_i^{in}, r^{out}, x_i', \bm{x}_{i-1}')
	=
	mean(\hat{\mathcal{X}}_i(r_i^{in}, r^{out}, x_i', \bm{x}_{i-1}')).
\end{equation}
\end{lemma}
For example, in Fig.~\ref{Fig:2User-example}, after observing $x_2'$ in the green dash line, the set of user $i$'s all possible real locations from the perspective of the adversary is $\hat{\mathcal{X}}_2=\{A,B,C,D\}$.
All these points are equally likely to appear given the symmetric query probability in Definition \ref{Definition:binary reporting}.

\subsubsection{User $i$'s Expected Privacy Gain under Definition \ref{Definition:binary reporting}}
Given the adversary's optimal inference in (\ref{Equ: ad infer}), we can determine user $i$'s expected privacy gain in (\ref{Equ: user i's expected privacy-average over x_i'}) under Definition \ref{Definition:binary reporting}.

\begin{prop}
\label{Prop:user i's privacy-appr}
By using the binary approximate query strategy of $x_i'$ in Definition \ref{Definition:binary reporting}, user $i$'s expected privacy gain in (\ref{Equ: user i's expected privacy-average over x_i'}) is given by: 
\begin{multline} 
\label{Equ: user i's pri under appr}
\pi_i^{appr}(r_i^{in},r^{out},\bm{x}_{i-1}')\\
=
\int_{x_i' \in \mathcal{M}} \frac{1}
{\int_{x_i' \in \mathcal{M}}|\hat{\mathcal{X}}_i(r_i^{in}, r^{out}, x_i', \bm{x}_{i-1}')|dx_i'} \\ 
\sum_{x_i \in \hat{\mathcal{X}}_i(r_i^{in}, r^{out}, x_i', \bm{x}_{i-1}')}  D(\hat{x}_i(r_i^{in}, r^{out}, x_i', \bm{x}_{i-1}'),x_i)dx_i'.
\end{multline}
Algorithm \ref{Algorithm: 1D}\footnote{The algorithm can be extended to an arbitrary distribution $\psi_i(x_i)$. The difference is that the adversary will no longer average among all the possible real locations 
but take a weighted mean based on $\psi_i(x_i)$.} returns the optimal obfuscation distance $r_i^{in}$ for Definition \ref{Definition:binary reporting} with computational complexity $\mathcal{O}(\frac{1}{\epsilon^2})$. 
\end{prop}

\begin{algorithm}[!t]
\label{Algorithm: 1D}
	\SetAlgoLined  
	\KwIn{${\bm{x}}_{i-1}'$, $Q_i$}
	\KwOut{$r_i^{in}$}
	
	Initialization: 
	 $r^{out}=\min(Q_i,\frac{1}{2})$,
	%
	%
	%
	%
$\mathcal{X}_i^{in}=\cup_{j=1}^{i-1} [x_j'-Q_i,x_j'+Q_i] \cap [0,1]$, 
$\hat{\mathcal{X}}_i=\emptyset$,
	
	
	\For{$r_i^{in}$=0:$\epsilon$:1} {

\For{$x_i'$=0:$\epsilon$:1     }{
			
Compute $\hat{\mathcal{X}}_i(r_i^{in}, r^{out}, x_i', \bm{x}_{i-1}')$ in (\ref{Equ: x_poss}) 
						
Compute $\hat{x}_i(x_i')
=
mean(\hat{\mathcal{X}}_i)$ in (\ref{Equ: ad infer})
			
Add elements $D(\hat{\mathcal{X}}_i,\hat{x}_i(x_i'))$ to $\tilde{\mathcal{D}}$   
}

$\pi_i^{appr}(r_i^{in}) = mean(\tilde{\mathcal{D}})$.
} 
	
$r_i^{in}=\arg\max \pi_i^{appr}(r_i^{in})$

\caption{Optimize $r_i^{in}
$ for any user $i$}  
\end{algorithm}  

So far, we have finished the optimal design of the binary approximate obfuscated scheme for any user $i$'s query strategy. Next we move on to the performance evaluation of the scheme returned by Algorithm \ref{Algorithm: 1D}.

\section{Evaluation of the Approximate Query}
\label{Sec:performance eva}

In this section, we first show our approximate scheme guarantees asymptotic optimum, as long as there are sufficient users in cooperation.
Then, we compare the approximate solution with the optimal but complicated cooperative strategy to problem (\ref{Opt:user 2's reporting prob-cont}) in Section \ref{Sec: pri coop against ad inference}, as well as the existing caching-based schemes in the literature \cite{zhang2019caching,shokri_hiding_2014,peng2017collaborative,hu2018proactive,jung2017collaborative,cui2020cache}.

\subsection{Asymptotic Optimum of Our Approximation Scheme} 
Lemma \ref{Lemma:X_k^{in}} shows that a finite number of users are already enough to cover all the PoIs for latter users. 
\begin{lemma} \label{Lemma:X_k^{in}}
There exists a finite user number $N'<\infty$ such that for any latter user $i\geq N'$, $\mathcal{X}_i^{in}(\bm{x}_{i-1}')=\mathcal{M}$. 
\end{lemma}

As long as we have enough users to cooperate, user $i \geq N'$ at a large enough order can always take advantage of former users' queries to meet the service constraints and hence achieve the maximum possible privacy gain.

\begin{prop}
\label{Prop:infinite N leads to opt privacy}
The expected privacy gain for an average user increases with the number $N$ of cooperative users.
As $N \rightarrow \infty$, our approximate obfuscated scheme is asymptotically optimal to solve problem (\ref{Opt:user 2's reporting prob-cont}). 
\end{prop}

\begin{figure}[!t]
\centering
\includegraphics[width=2.4 in]{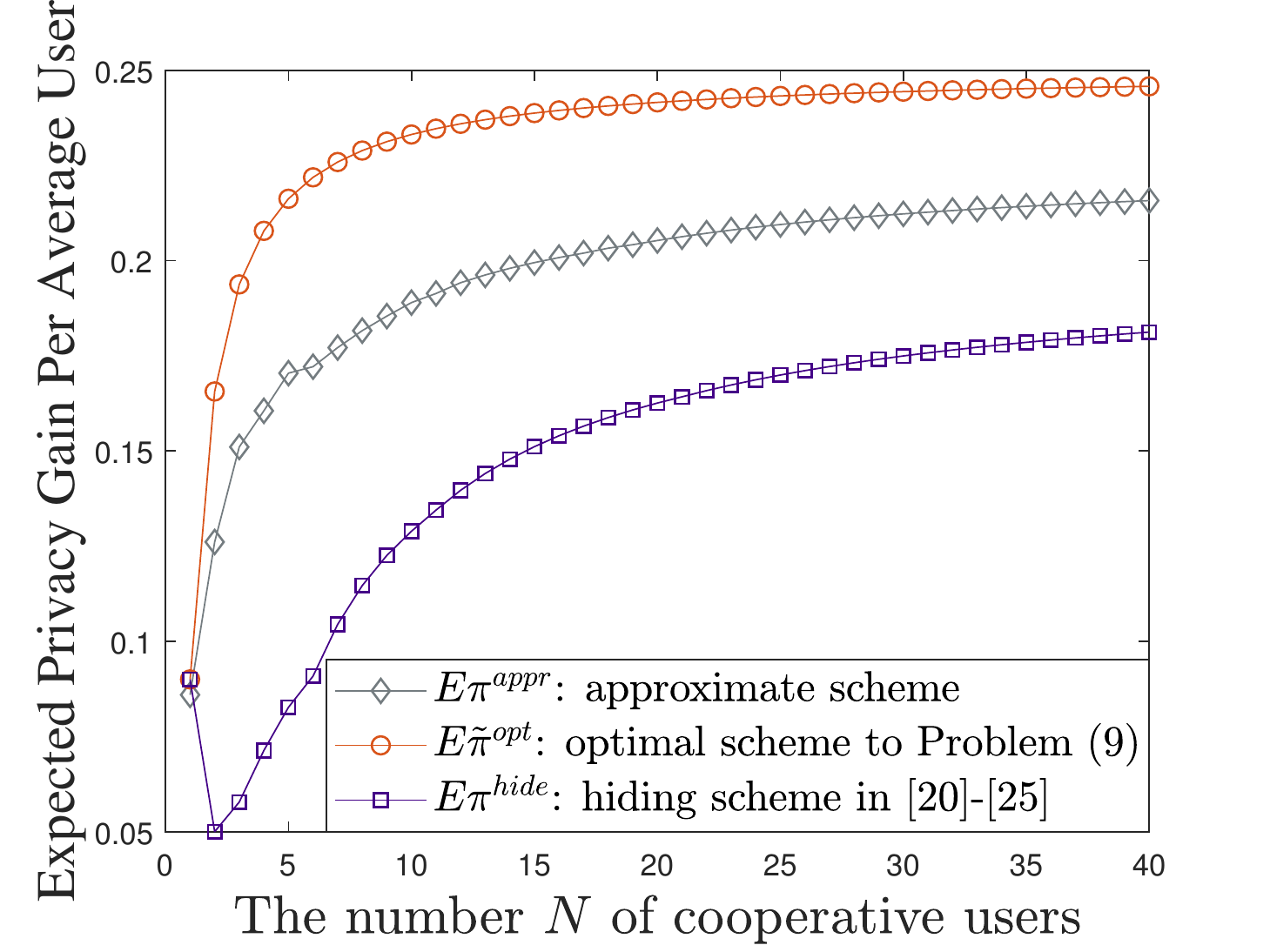}
\caption{Performance evaluation of the approximate obfuscated query scheme ($\mathbb{E}\pi^{appr}$) with the optimal strategy ($\mathbb{E}{\tilde{\pi}}^{opt}$) and hiding schemes in the literature \cite{zhang2019caching,shokri_hiding_2014,peng2017collaborative,hu2018proactive,jung2017collaborative,cui2020cache}. 
Users' $Q_i$ follows the i.i.d. truncated normal distribution with mean $\mu=0.1$ and standard deviation $\sigma=0.1$.} 
\label{Fig:compare with opt}
\end{figure}

Proposition \ref{Prop:infinite N leads to opt privacy} shows the privacy improvement of our approximate scheme thanks to a great number of users to cooperate and share PoI information.

\subsection{Approximation Ratio under a Finite User Number}

Besides examining the performance for a large user number, we also compare our approximate obfuscated query scheme in Section \ref{Sec:approximation1D} with the optimal solution to Problem (\ref{Opt:user 2's reporting prob-cont}) 
in Section \ref{Sec: pri coop against ad inference}. 

As Problem (\ref{Opt:user 2's reporting prob-cont}) can only be numerically solved without any analytical expression of the optimal performance, we replace with the upper bound of the optimal solution, i.e., $\pi_i^{opt} \leq \frac{1}{4}$ from Proposition \ref{Prop: user 1 strategy}.

\begin{prop}
\label{Prop: performeance of appr}
Our approximate query strategy returned by Algorithm \ref{Algorithm: 1D} reaches at least $\frac{3}{5}$ of the maximum privacy gain in Problem (\ref{Opt:user 2's reporting prob-cont}).
\end{prop}
Note that this approximation ratio is a loose bound for ease of analysis, and the actual performance ratio is much better.

\subsection{Simulations for Performance Comparison}

Fig.~\ref{Fig:compare with opt} presents a numerical example for the expected privacy gain $\mathbb{E}\pi^{appr}$ of the approximate obfuscated query scheme for an average user.
As $N$ increases, $\mathbb{E}\pi^{appr}$ approaches the maximum possible privacy gain $\frac{1}{4}$, yet with a diminishing return.
It shows that our approximate scheme has limited gap with optimum, yet it saves much time for computing. This gap reduces as we have more users to cooperate, which is consistent with Proposition \ref{Prop:infinite N leads to opt privacy} .

Recall in Subsection \ref{Subsec: hiding compared with opt}, we follow the hiding-from-the-LBS idea in the literature to compare with the optimal cooperative strategy. 
For a fair comparison, here we consider the benchmark case: a user only queries with the binary approximate query strategy when he does not find useful information in the cache and simply hides from the LBS (no query) otherwise as in \cite{zhang2019caching,shokri_hiding_2014,peng2017collaborative,hu2018proactive,jung2017collaborative,cui2020cache}.
We can observe that our approximation scheme obviously outperforms the hiding scheme, to show the advantage of always querying strategically.

\section{Optimal Query Sequence among Users}
\label{Sec:order}
Besides guiding LBS user's cooperative query strategy,
%
when users come with heterogeneous service flexibility $Q_i$, we can still optimize the query sequence of users.
In this section, we study the optimal query sequence to enhance the multi-user privacy protection  performance.
Notice that user set $\mathcal{N}$ no longer follows sequence $\{1,\cdots,N\}$, and the total expected privacy gain for $N$ users is given as
\begin{align}
\label{Equ:total privacy Pi}
\Pi(\mathcal{N})
&= 
\sum_{i \in \mathcal{N}}\pi_i.
\end{align}

First we consider the simple but fundamental scenario with only two users with $Q_1\leq Q_2$.
Recall that in Sections \ref{Subsec:first user} and \ref{Subsec:user 2 coop 1D appr}, we analytically solve the approximate query strategy for the users at the first and the second order, and obtain the corresponding privacy gains.
To compare two different query sequences to have user 1 or 2 first to query for maximizing (\ref{Equ:total privacy Pi}), we obtain the optimal query sequence.

\begin{prop}
\label{Prop: order-2 small}
When both users have small service flexibilities (i.e., $0\leq Q_1 \leq Q_2 \leq \frac{1}{11}$),
the optimal sequence to maximize the total expected privacy gain is to let user 1 with the smaller service flexibility $Q_1$ query the LBS platform first.
\end{prop}

The intuition behind is that when there are two users with relatively tight service constraints, user 1 with a smaller service flexibility $Q_1$ can still hardly benefit much from other queries. 
Yet letting user 1 query first provides a greater service coverage region $\mathcal{X}_2^{in}(x_1')$ to preserve user 2's privacy.

\begin{prop}
\label{Prop: order-1 small 1 large}
When one user has a small service flexibility while the other has a large service flexibility (i.e., $0\leq Q_1 \leq  \frac{1}{11}$ and $Q_2 \rightarrow \frac{1}{2}$), the optimal sequence to maximize the total expected privacy gain is to let user 2 with the greater service flexibility $Q_2$ query the LBS platform first.
	
\end{prop}
When user 2 have a greater service flexibility $Q_2$, his privacy gain already approaches the maximum possible value and can hardly be improved from the cooperation. Thus in this case, we let user 2 query first to benefit latter user 1 with a tighter service constraint.
This opposite result from Proposition \ref{Prop: order-2 small} implies that the optimal query sequence depends on the exact value of users' service flexibilities $Q_i$ and shows no simple monotonicity.

Using the simulation, we can extend such the insights from Propositions \ref{Prop: order-2 small} and \ref{Prop: order-1 small 1 large} to a more general setting.
\begin{obs}[Optimal query sequence for $N=2$]
\label{Obs:2user}
When $Q_1=0.1$ is fixed, there exists two thresholds $0.15$ and $0.3$ for the value of $Q_2$ such that 
\begin{itemize}
\item when $0.15<Q_2 < 0.3$, the optimal sequence to maximize the total privacy gain of the two users is $\mathcal{N}=\{1,2\}$, which extends the insight from Proposition \ref{Prop: order-2 small}.

\item when $Q_2 \leq 0.15 $ or $Q_2 \geq 0.3$, the optimal sequence to maximize the total privacy gain of the two users is $\mathcal{N}=\{2,1\}$, which extends the insight from Proposition \ref{Prop: order-1 small 1 large}. 

\end{itemize}

\end{obs}

\begin{figure}[!t]
\centering
\includegraphics[width=2.4 in]{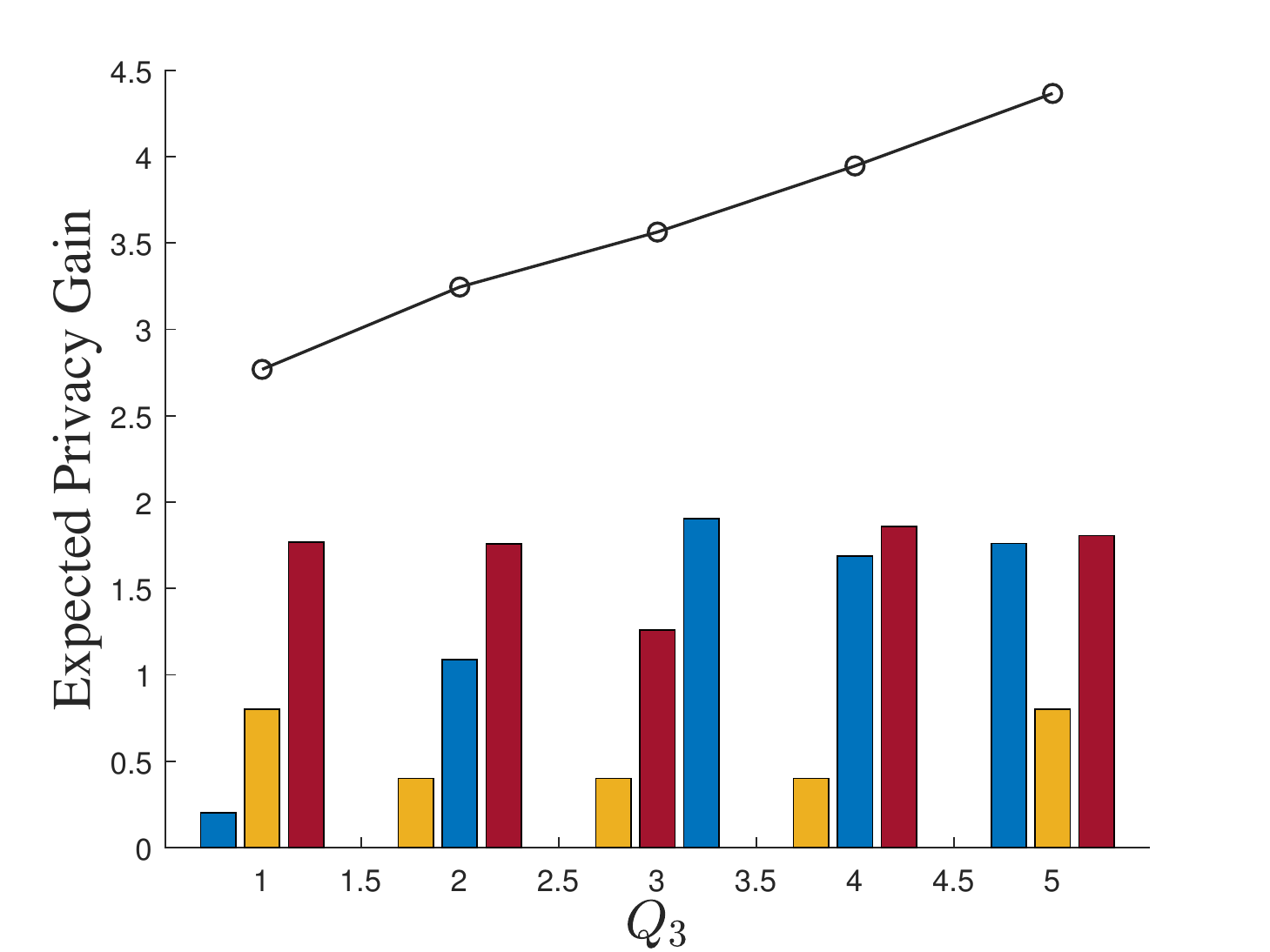}
\caption{The optimal sequence for three-user cooperation versus $Q_3$ when fixing $Q_1=0.1$ and $Q_2=0.2$.}
\label{Fig:User3Order}
\end{figure}

To see whether Observation \ref{Obs:2user} also applies to more than two users, we numerically study on three users' cooperation and obtain the optimal query sequence leading to the maximal total privacy gain $\Pi$ in Fig.~\ref{Fig:User3Order}. We vary user 3's service flexibility $Q_3$ while fixing $Q_1=0.1$ and $Q_2=0.2$.
The total expected privacy gain $\Pi$ increases with $Q_3$.

Given users 1 and 2 have similarly small service flexibilities as in Proposition \ref{Prop: order-2 small}, we ask user 1 to query before user 2. If $Q_3$ is similarly small (i.e., $0 \leq Q_3<0.25$), we assign user 3 to be the first, second and third optimally as $Q_3$ increases (see the blue bar). 
As $Q_3$ becomes greater than $Q_1$ and $Q_2$ (i.e., $Q_3>0.19$), we assign user 3 to be the third, second and first optimally as $Q_3$ increases.
The insight behind is that if user 3's service flexibility $Q_3$ takes extreme values (i.e., too large or too small), his privacy improvement shows little sensitivity to the query sequence, then he should query the LBS platform first. If $Q_3$ is relatively in the middle of its value range (i.e., $0.19 <Q_3<0.25$), then the user should be the last to query, as his privacy improvement is more sensitive to the query sequence. Similar results can be found for greater user numbers $N$ and the numerical thresholds can be used to guide the crowdsourcing system for a better total privacy performance.

\section{Extension to 2D Location Scenario}
\label{Subsec:extension 2D}
Recall in Section \ref{Sec:approximation1D}, we give a binary approximate obfuscated query scheme to design the user's query strategy in a 1D line interval $\mathcal{M}=[0,1]$ with only left- and right-hand side directions. 
In this section, we extend the approximate query strategy to the normalized 2D ground plane (i.e., $\mathcal{M}=[0,1]\times [0,1]$) using similar design and analysis. For user $i$ with real location $(x_i,y_i)$, we extend the approximate obfuscated query in four different directions (north, south, east and west)  with equal obfuscation distances.

\begin{algorithm}[!t]
\label{Algorithm: 2D}
\SetAlgoLined  
\KwIn{$({\bm{x}}_{i-1}', {\bm{y}}_{i-1}')$, $Q_i$}
\KwOut{$r_i^{in}$}

Initialization: 
$r^{out}=\min(Q_i,\frac{1}{2})$,
$\mathcal{M}=[0,1]\times[0,1]$,
$\mathcal{X}_i^{in}
=\cup_{j=1}^{i-1}
\{(x_i,y_i) \in \mathcal{M} |
D((x_i,y_i),(x_i',y_i')) \leq Q_i
\}
\cap \mathcal{M}$,
$\hat{\mathcal{X}}_i=\emptyset$

\For{$r_i^{in}$=0:$\epsilon$:1} {
	
\For{$x_i'$=0:$\epsilon$:1     }{

Compute $\hat{\mathcal{X}}_i(x_i')$ in (\ref{Equ: x_poss}) 
						
Compute $\hat{x}_i(x_i')
=
mean(\hat{\mathcal{X}}_i(x_i'))$ in (\ref{Equ: ad infer})
		
\For{$y_i'$=0:$\epsilon$:1     }
		{
			
Compute $\hat{\mathcal{Y}}_i(y_i')$ in (\ref{Equ: x_poss}) 
						
Compute $\hat{y}_i(y_i')
=
mean(\hat{\mathcal{Y}}_i(y_i'))$ in (\ref{Equ: ad infer})
			
Add elements $D((\hat{\mathcal{X}}_i,\hat{\mathcal{Y}}_i),(\hat{x}_i,\hat{y}_i))$ to $\tilde{\mathcal{D}}$   
		}
	}

$\pi_i^{appr}(r_i^{in})=mean(\tilde{\mathcal{D}})$

} 

$r_i^{in}=\arg\max \pi_i^{appr}(r_i^{in})$

\caption{Optimize $r_i^{in}$ under the four-point approximation for any user $i$ in 2D}  
\end{algorithm}  

\begin{defn}[Four-point approximate obfuscation for LBS query]
\label{Definition:four- reporting}
In the 2D domain,  if a user $i$ finds the cached PoI information useful, i.e.,
\begin{align*}
&(x_i,y_i)\in \mathcal{X}_i^{in}(\bm{x'}_{i-1},\bm{y'}_{i-1}) \\
=&\cup_{j=1}^{i-1}
\{(x_i,y_i) \in \mathcal{M} |
D((x_i,y_i),(x_j',y_j')) \leq Q_i
\},
\end{align*}
his strategy $f_i^{in}$ is to randomize the query among four points with a fixed obfuscation distance $r_i^{in}$ away at the four sides of his real location $(x_i,y_i)$:
\begin{equation*}
\label{Equ: f_in qua def 2D}
\begin{aligned}
&f_i^{in}((x_i',y_i')|(x_i,y_i),(\bm{x'}_{i-1},\bm{y'}_{i-1})) \\ = 
&\begin{cases} 
\frac{1}{4}, \text{ if } 
(x_i',y_i')=(|x_i-r_i^{in}|,|y_i-r_i^{in}|), \\
\frac{1}{4}, \text{ if } 
(x_i',y_i')=(|x_i-r_i^{in}|,1-|y_i-1+r_i^{in}|), \\
\frac{1}{4}, \text{ if } 
(x_i',y_i')=(1-|x_i-1+r_i^{in}|,|y_i-r_i^{in}|), \\
\frac{1}{4}, \text{ if } 
(x_i',y_i')=(1-|x_i-1+r_i^{in}|,1-|y_i-1+r_i^{in}|).
	\end{cases} 
\end{aligned}
\end{equation*}
Otherwise if $(x_i,y_i)\in \mathcal{X}_i^{out}(\bm{x'}_{i-1},\bm{y'}_{i-1})=\mathcal{M} \setminus \mathcal{X}_i^{in}$, 
\begin{equation*}
\label{Equ: f_out qua def 2D}
\begin{aligned}
&f_i^{out}((x_i',y_i')|(x_i,y_i),(\bm{x'}_{i-1},\bm{y'}_{i-1})) \\ = 
&\begin{cases} 
\frac{1}{4}, \text{ if } 
(x_i',y_i')=(|x_i-r_i^{out}|,|y_i-r^{out}|), \\
\frac{1}{4}, \text{ if } 
(x_i',y_i')=(|x_i-r_i^{out}|,1-|y_i-1+r^{out}|), \\
\frac{1}{4}, \text{ if } 
(x_i',y_i')=(1-|x_i-1+r_i^{out}|,|y_i-r^{out}|), \\
\frac{1}{4}, \text{ if } 
(x_i',y_i')=(1-|x_i-1+r_i^{out}|,1-|y_i-1+r^{out}|).
	\end{cases} 
\end{aligned}
\end{equation*}
\end{defn}
In the $x-$ and $y-$domain, we can consider the obfuscated query $f_i((x_i',y_i')|(x_i,y_i),(\bm{x'}_{i-1},\bm{y'}_{i-1}))$ separately for $x_i$ and $y_i$. Given the reduction from 2D to 1D, we can apply Definition \ref{Definition:binary reporting} for $f_i(x_i'|x_i,\bm{x'}_{i-1},Q_i)$ and $f_i(y_i'|y_i,\bm{y'}_{i-1})$ in the 1D domain. Then the adversary's inference is consisted of $\hat{x}_i(x_i'|\bm{x'}_{i-1})$ and $\hat{y}_i(y_i'|\bm{y'}_{i-1})$ the same as in 1D.
Therefore, user $i$'s expected privacy gain in (\ref{Equ: user i's expected privacy-average over x_i'}) can be given in 2D as
\begin{multline}
\label{Equ:user i's  privacy 2D}
\pi_i^{appr}=
\int_{(x_i',y_i') \in \mathcal{M}}
\operatorname{Pr} ((x_i',y_i')|(\bm{x'}_{i-1},\bm{y'}_{i-1})) \\
\min_{(\hat{x}_i,\hat{y}_i)\in \mathcal{M}}  
\int_{(x_i,y_i) \in \mathcal{M}} \operatorname{Pr}((x_i,y_i)|(x_i',y_i'),(\bm{x'}_{i-1},\bm{y'}_{i-1})) \\
D((\hat{x}_i,\hat{y}_i),(x_i,y_i)) dx_i dy_i
dx_i' dy_i',
\end{multline}
where $(\hat{x}_i,\hat{y}_i)$ is the adversary's inference attack similarly from (\ref{Equ:inference error for user 2}). 
Similarly to Proposition \ref{Prop:user i's privacy-appr}, we can rewrite the user's privacy gain under the four-point obfuscation.
\begin{prop}
\label{Prop:user i's privacy-4 appr}
By using a four-point approximate obfuscated query strategy in 2D in Definition \ref{Definition:four- reporting}, user $i$'s expected privacy gain in (\ref{Equ:user i's  privacy 2D}) can be rewritten by
\begin{multline} 
\label{Equ: user i's pri under 4 appr}
\pi_i^{appr}(r_i^{in},r^{out},(\bm{x'}_{i-1},\bm{y'}_{i-1}))\\
=\int_{(x_i',y_i')\in \mathcal{M}} 
	\frac{1}
{\sum_{x_i'}
|\hat{\mathcal{X}}_i|\times
\sum_{y_i'}|\hat{\mathcal{Y}}_i}|
	\sum_{x_i \in \hat{\mathcal{X}}_i}
	\sum_{y_i \in \hat{\mathcal{Y}}_i} \\
D((\hat{x}_i,\hat{y}_i),(x_i,y_i)) dx_i' dy_i',
\end{multline}
where $\hat{\mathcal{X}}_i=\hat{\mathcal{X}}_i(r_i^{in},r^{out},x_i',\bm{x}_{i-1}')$ and $\hat{\mathcal{Y}}_i=\hat{\mathcal{Y}}_i(r_i^{in},r^{out},y_i',\bm{y}_{i-1}')$ can be obtained by solving equations from (\ref{Equ: x_poss}).

Algorithm \ref{Algorithm: 2D} returns the optimal obfuscation distance $r_i^{in}$ for Definition \ref{Definition:binary reporting} with computational complexity $\mathcal{O}(\frac{1}{\epsilon^3})$. 
\end{prop}

Compared with Algorithm \ref{Algorithm: 1D} in 1D, Algorithm \ref{Algorithm: 2D} iterates similarly for the adversary's inference attack in one more dimension.
To provide more practical performance evaluation of our approach, we use the datasets of real-world map \cite{NYCmap} and search interests for the term `restaurant' from Google Trends \cite{SearchData} in New York city for different time slots in a day.
Under this general 2D ground plane scenario, users' locations are no longer uniformly distributed, and thus we next refine the adversary's inference procedure in Algorithm \ref{Algorithm: 2D}.
As the real location $x_i$ is not uniformly distributed, the candidates in sets $\hat{\mathcal{X}}_i(x_i')$ and $\hat{\mathcal{Y}}_i(y_i')$  for the real location are no longer inferred with equal probabilities. Therefore, we modify Step 5 in Algorithm \ref{Algorithm: 2D} as
\begin{equation*}
\text{Compute } \hat{x}_i(x_i')
=
\sum_{x \in \hat{\mathcal{X}}_i(x_i')}
\frac{\psi_i^X(x) \cdot x}
{
\sum_{x \in \hat{\mathcal{X}}_i(x_i')}\psi_i^X(x)
},
\end{equation*}
where $\psi_i^X(x)=\sum_{y\in [0,1]}\psi_i(x,y_i)$ is user $i$'s general location distribution in the $x-$domain of the 2D plane.
Step 8 in the $y-$domain can be modified similarly.
For the user $i$'s expected privacy gain in Steps 9 and 12, we modify the calculation as

$$
\pi_i^{appr}(r_i^{in})=
\sum_{x \in \hat{\mathcal{X}}_i(x_i')}
\sum_{y \in \hat{\mathcal{Y}}_i(y_i')}
\frac{\psi_i(x,y)
D((x,y),(\hat{x}_i,\hat{y}_i))
}
{
\sum_{x \in \hat{\mathcal{X}}_i(x_i')}
\sum_{y \in \hat{\mathcal{Y}}_i(y_i')}
\psi_i(x,y)
}.
$$

\begin{figure}[!t]
\centering
\subfigure[Number of searches for restarurants in New York city in different time slots.]{
\includegraphics[width=1.65 in]{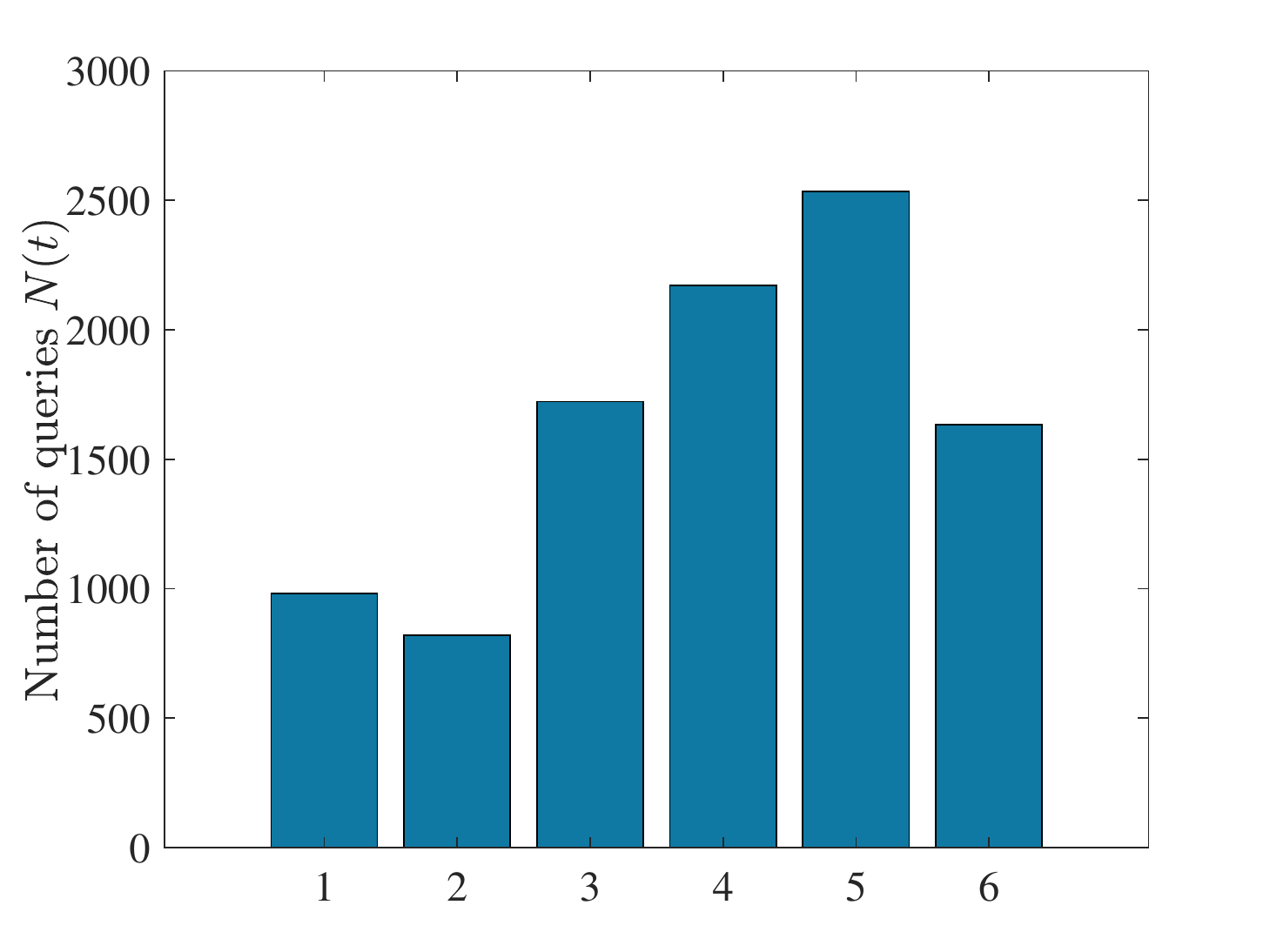}
\label{Subfig:number of searches}
}
\subfigure[Expected privacy gain for an average user searching for restaurants. In different time slots, numbers of cooperative users follow $N(t)$ in Fig.~\ref{Subfig:number of searches} and users' location distributions $\psi_i$ are i.i.d. following the real-life map in New York city. ]{
\includegraphics[width=1.65 in]{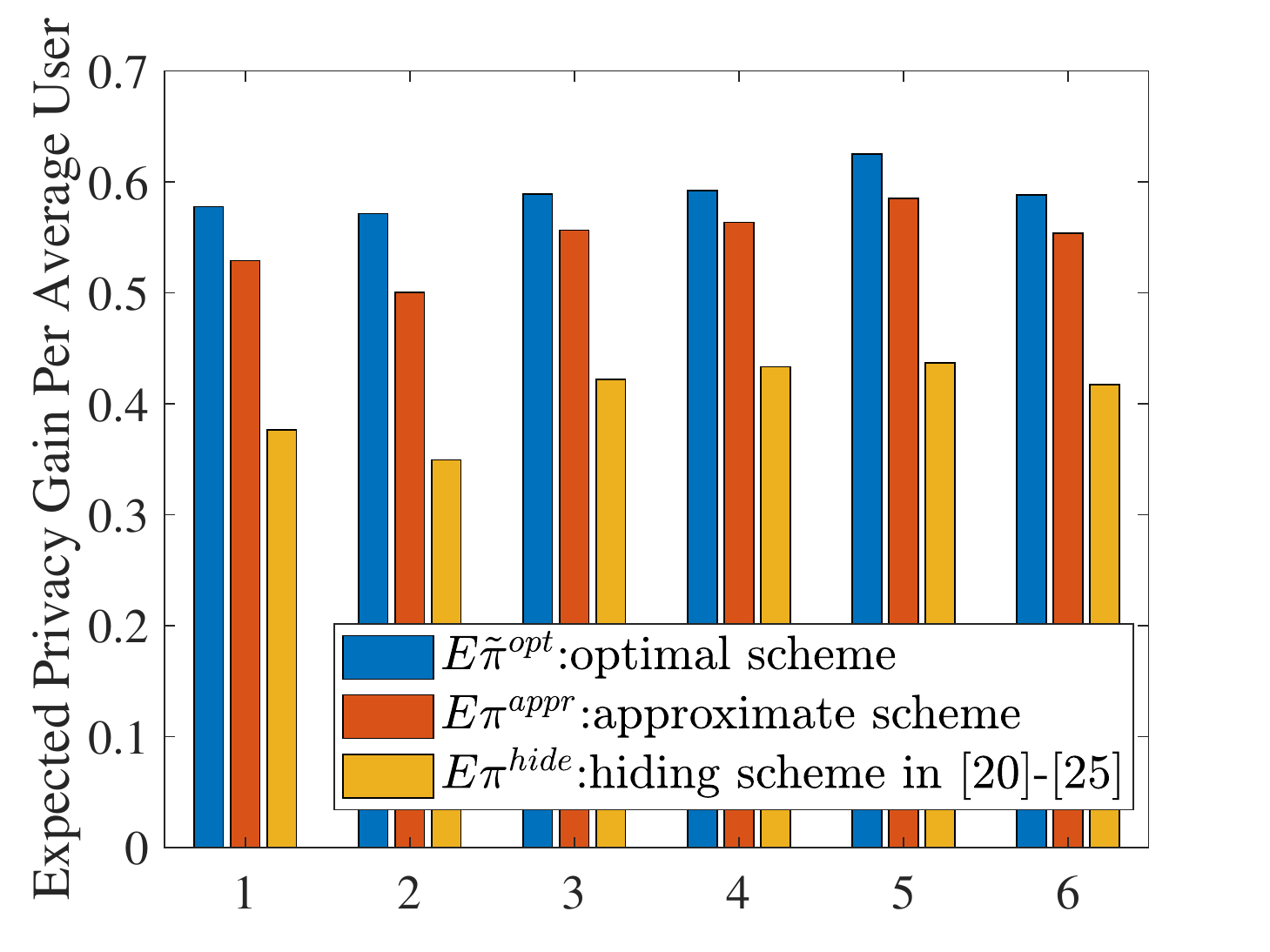}
\label{Subfig:privacy 2D NY}
}
\caption{
Performance evaluation of our (four-point) approximate obfuscated query scheme ($\mathbb{E}\pi^{appr}$) in 2D from Algorithm \ref{Algorithm: 2D} with our optimal strategy ($\mathbb{E}{\tilde{\pi}}^{opt}$) and the hiding scheme in the literature \cite{zhang2019caching,shokri_hiding_2014,peng2017collaborative,hu2018proactive,jung2017collaborative,cui2020cache}. 
Here, any user's service flexibility degree $Q_i$ follows the i.i.d. truncated normal distribution with mean $\mu=0.1$ and standard deviation $\sigma=0.1$.
}
\label{Fig:}
\end{figure}

Based on the dataset on Google Trends \cite{SearchData}, Fig.~\ref{Subfig:number of searches} tells the number of searches for one kind of PoI (restaurant). 
Following the real-world population map in New York city \cite{NYCmap}, we further generate an i.i.d. location distribution $\psi_i$ for all cooperative users.
Fig.~\ref{Subfig:privacy 2D NY} then gives the comparison of $\mathbb{E}{\tilde{\pi}}^{opt}$, $\mathbb{E} \pi^{appr}$ and $\mathbb{E} \pi^{hide}$.
Our approximate obfuscated query scheme is obviously better than the hiding scheme in the literature, and is close to optimal solution which suffers from high computational complexity. 

\section{Conclusion}
\label{Sec:conclusion}
To address potential data leakage from LBS platforms as well as peers, we study on multi-user privacy preservation problem under a cooperation scheme. 
We propose to leverage LBS users' service flexibility with one single query for a better privacy gain.
Depending on whether a user benefits from the crowdsourced cache, he takes different query strategies with obfuscation in a distributed way and contribute to the cache to benefit latter users. 
Differently from the existing literature, we recommend users to query with obfuscated locations to the LBS platform even if they find useful PoI information in the cache, to jointly increase confusion to the adversary and store more PoI data.
To relax the complexity, we simplify the query scheme to a two-point randomized query in 1D and a four-point randomized query in 2D. Both of them show guaranteed performance compared with the optimal scheme and significant privacy improvement compared with hiding schemes in the literature.
From the perspective of the crowdsourced system, we also study the optimal query sequence to maximize the total expected privacy gain.

\bibliographystyle{ieeetr}

\bibliography{TMCRef}

%

\newpage
\appendices
\onecolumn

\section{Proof of Proposition \ref{Prop:prob reformulation-discrete}}
\label{Proof:Prop:prob reformulation-discrete}
\begin{IEEEproof}
Let
\begin{align*}
y_{ad}=
\min_{\hat{x}_i} 
\left( 
\sum_{x_i \in \mathcal{X}_i^{in}} \psi_i(x_i) f_i^{in}(x_i'|x_i) D(\hat{x}_i,x_i) \right. \\
+
\left. \sum_{x_i \in \mathcal{X}_i^{out}} \psi_i(x_i) f_i^{out}(x_i'|x_i) D(\hat{x}_i,x_i)\right)
\end{align*}
then
Problem (\ref{Opt:user 2's reporting prob-discrete}) can be rewritten with constraints as an LP problem as
\begin{multline}
\max
\sum_{x_i'}
y_{x_i'}  \\
\text{s.t. }
y_{x_i'} \leq 
\sum_{x_i \in \mathcal{X}_i^{in}} \psi_i(x_i) f_i^{in}(x_i'|x_i) D(\hat{x}_i,x_i)  \\
+
\sum_{x_i \in \mathcal{X}_i^{out}} \psi_i(x_i) f_i^{out}(x_i'|x_i) D(\hat{x}_i,x_i), \forall x_i',\hat{x}_i \in \mathcal{M}, \\
f_i^{out}(x_i'|x_i)=0,  \\
\forall (x_i, x_i') \in 
\left\lbrace(x_i, x_i')| 
x_i\in \mathcal{X}_i^{out}, x_i'\in \mathcal{M}: D(x_i',x_i)> Q_i \right\rbrace,\\ 
\text{var:} f_i^{in}(x_i'|x_i), f_i^{out}(x_i'|x_i), \forall x_i,x_i' \in \mathcal{M},
\end{multline} 
which follows the idea of \cite{shokri2016privacy}. 

Given that Karmarkar \cite{karmarkar1984new} proved that an LP problem with $n$ variables and $m$ constraints can be solved by an interior algorithm within complexity $\mathcal{O}(m^{3/2}n^2)$.
The LP problem above has $M+M^2$ variables and $M^2+M+1$ constraints, 
where $M$ is the number of the finite regions for users' locations, thus the time complexity for the user's privacy protection problem is $\mathcal{O}( M^{7})$.
\end{IEEEproof}

\section{Proof of Lemma \ref{lemma:hide}}
\label{Proof:lemma:hide}
To prove Lemma \ref{lemma:hide}, first we have the following lemma.
\begin{lemma}
For any user at order $i$, $\pi_i^{opt} \geq \pi_i^{hide}$ always holds, and the equality only holds if $\mathcal{X}_i^{in}=\emptyset$ or $\mathcal{X}_i^{in}=\mathcal{M}$.
\end{lemma}

\begin{IEEEproof}
Considering the hiding-from-the-LBS scheme from the literature when a user already finds useful information in the cache, it is equivalent to $f_i^{in}=\emptyset$ in Problem (\ref{Opt:user 2's reporting prob-cont}). In this case, the adversary's inference when it observes no query is to randomly infer one location from the covered location set $\mathcal{X}_i^{in}(\bm{x}_{i-1}')$.

Let $\mathcal{M}^{out}$ denote the queried location set for all $x_i \in \mathcal{X}_i^{out}$, i.e.,
\begin{equation*}
\mathcal{M}^{out}=\{x\in \mathcal{M}| f_i^{out}(x|x_i)> 0\}.
\end{equation*}
Then let a reported query $x_i' \in \mathcal{M} \setminus \mathcal{M}^{out}$, user $i$'s privacy is equivalent to that with $f_i^{in}=\emptyset$, as the adversary will similarly infer from the location set corresponding to the query $x_i'$.
However, if $x_i' \in \mathcal{M}^{out}$, the randomization increases when the adversary observes the query $x_i'$. Specifically, let a query $x_i'=z' \in \mathcal{M}^{out}$, and 
\begin{equation*}
\mathcal{Z}^{out}=\{x\in \mathcal{M}|f_i^{out}(z|x_i)> 0\}
\end{equation*}
denotes the real location set at which user $i$ queries with $z$.
Under the hiding scheme with $f_i^{in}=\emptyset$, the adversary's inference is randomized among the set $\mathcal{Z}^{out}$, i.e., $\hat{x}_i=mean(\mathcal{Z}^{out})$.
Yet under our always-query scheme with a given strategy $f_i^{in}(z|x_i)=1$ for any $x_i \in \mathcal{X}_i^{in}$, the randomization range for the adversary is enlarged to $\mathcal{Z}^{out} \cup \mathcal{X}_i^{in}$ and $\hat{x}_i=mean(\mathcal{Z}^{out} \cup \mathcal{X}_i^{in})$, which leads to a better privacy gain for both $x_i \in \mathcal{Z}^{out} $ and $x_i \in  \mathcal{X}_i^{in}$.
Notice that this does not leads to our optimal privacy gain $\pi_i^{opt}$ but provides a lower bound performance under our always-query scheme. 
This completes our proof of $\pi_i^{opt} > \pi_i^{hide}$.

When $\mathcal{X}_i^{in}=\emptyset$, hiding scheme does not exist. When $\mathcal{X}_i^{in}=\mathcal{M}$, there will be no left location set for our $f_i^{in}$ design. There we have $\pi_i^{opt}= \pi_i^{hide}$ and complete the proof.
\end{IEEEproof}

Given a user at any order $i$ has $\pi_i^{opt} \geq \pi_i^{hide}$ and the equality does not always hold, in the expected sense, we have $\mathbb{E}\pi^{opt} \geq \mathbb{E}\pi^{hide}$.

\section{Proof of Proposition \ref{Prop: user 1 strategy}}
\label{Proof:Prop: user 1 strategy}
\begin{IEEEproof}
To prove Proposition \ref{Prop: user 1 strategy}, we first have the following Lemma:
\begin{lemma}
\label{Lemma:max privacy}
The first user 1’s maximum possible expected privacy gain without any service constraint is $\frac{1}{4}$.
\end{lemma}
\begin{IEEEproof}
For the first user with $ \mathcal{X}^{in}=\emptyset$, the objective in (\ref{Prop:prob reformulation-discrete}) becomes
\begin{multline}
\max
\sum_{x_1'}
\min_{\hat{x}_1} 
\left( 
\sum_{x_1 \in \mathcal{M}} \psi_1(x_1) f_1^{out}(x_1'|x_1,\emptyset) D(\hat{x}_1,x_1)\right) 
\end{multline} 
and the corresponding LP problem becomes
\begin{equation}
\label{Opt: LP of first user}
\begin{aligned}
\max_{f_1^{out}(x_1'|x_1,\emptyset),y_{x_1'}} &\sum_{x_1'} y_{x_1'} \\
\text{s.t. }
&y_{x_1'} \leq \sum_{x_1 \in \mathcal{M}} \psi_1(x_1) f_1^{out}(x_1'|x_1,\emptyset) D(\hat{x}_1,x_1), \forall x_1',\hat{x}_1, \\
&\sum_{x_1 \in \mathcal{M}} \psi_1(x_1) f_1^{out}(x_1'|x_1,\emptyset) D(\hat{x}_1,x_1) \leq Q_{1}, \\
&\sum_{x_1'} f_1^{out}(x_1'|x_1,\emptyset)=1, \forall x_1, \\
&f_1^{out}(x_1'|x_1,\emptyset) \geq 0, \forall x_1, x_1'.
\end{aligned}
\end{equation}
We have the following Proposition.
\begin{prop} \label{Prop:IPP increases with Q^max and unchanged}
The objective function in the LP problem (\ref{Opt: LP of first user}) is non-decreasing with $Q_1$, and stays at the maximum privacy level after a threshold $Q_1^{MAX}$ for $Q_1$.
Given the user's prior information $\psi_1(x_1)$ and privacy measure $D(\cdot)$, the upper bound of the privacy can be reached for a large enough $Q_1$ when the user misreports according to the prior location distribution, i.e., $f_1^{out}(x_1'|x_1,\emptyset)=\psi_1(x_1')$.
\end{prop}
\begin{IEEEproof}
Notice  that as $Q_1$ increases, the service quality constraint becomes looser, thus the objective function is non-decreasing as $Q_1$ increases.  
When $Q_1$ goes to infinity, the quality constraint can be ignored and the objective value to be optimized can be obtained by solving
\begin{multline}
\label{Opt:IPP-quality relaxation}
\max_{f_1^{out}(x_1'|x_1,\emptyset),y_{x_1'}} \sum_{x_1'} y_{x_1'} \\
\text{s.t. }
y_{x_1'} \leq \sum_{x_1 \in \mathcal{M}} \psi_1(x_1) f_1^{out}(x_1'|x_1,\emptyset) D(\hat{x}_1,x_1), \forall x_1',\hat{x}_1, \\
\sum_{x_1'} f_1^{out}(x_1'|x_1,\emptyset)=1, \forall x_1, \\
f_1^{out}(x_1'|x_1,\emptyset) \geq 0, \forall x_1, x_1',
\end{multline}
which is a limited objective value as
\begin{align*}
\sum_{x_1'} y_{x_1'}
&\leq 
\sum_{x_1'} \sum_{x_1} \psi_1(x_1) f_1^{out}(x_1'|x_1,\emptyset) D(\hat{x}_1,x_1) \\
&\leq 
\sum_{x_1} \psi_1(x_1) D(\hat{x}_1,x_1) .
\end{align*}
The optimal misreport strategy $f_1^*(x_1'|x_1)$ obtained from Problem (\ref{Opt:IPP-quality relaxation}) is the maximum possible privacy level for the user without any quality constraint, and we have the corresponding quality loss as 
$$\sum_{x_1 \in \mathcal{M}} \psi_1(x_1) f_1^*(x_1'|x_1) D(\hat{x}_1,x_1)=Q_1^{MAX}.$$
Then for any $Q_1 \geq Q_1^{MAX}$, the maximum privacy level can be achieved and the objective function for Problem (\ref{Opt: LP of first user}) stays the same.

Next we prove the LP objective's upper bound.
With a large enough $Q_1$, i.e., $Q_1 \geq Q_1^{MAX}$, the objective is equivalent to
\begin{equation}
\begin{aligned}
&\sum_{x_1'} \min_{\hat{x}_1} \sum_{x_1} f(x_1'|x_1) \psi_1(x_1)d_p (\hat{x}_1,x_1) \\
&\leq
\min_{\hat{x}_1} \sum_{x_1'} \sum_{x_1} f(x_1'|x_1) \psi_1(x_1)d_p (\hat{x}_1,x_1) \\
&=
\min_{\hat{x}_1} \sum_{x_1} \sum_{x_1'} f(x_1'|x_1) \psi_1(x_1)d_p (\hat{x}_1,x_1) \\
&=
\min_{\hat{x}_1} \sum_{x_1} \psi(x_1)d_p (\hat{x}_1,x_1),
\end{aligned}
\end{equation}
where the first inequality is due to
\begin{equation}
\sum_y \min_x f(x,y) 
\leq \sum_y f(x^*,y) 
= \min_x \sum_y f(x,y),
\end{equation}
and the second equality is due to  $\sum_{x_1'} f(x_1'|x_1)=1$.
Notice that $\min_{\hat{x}_1} \sum_r \psi_1(x_1)d_p (\hat{r},x_1)$ is a fixed constant given $\psi_1(x_1)$ and $d_p(\cdot)$. 
When we set $f(x_1'|x_1)=\psi(x_1')$, the objective function becomes
\begin{equation}
\begin{aligned}
&\sum_{x_1'} \min_{\hat{x}_1} \sum_{x_1} f(x_1'|x_1) \psi_1(x_1)d_p (\hat{x}_1,x_1) \\
&=
\sum_{x_1'} \min_{\hat{x}_1} \sum_{x_1} \psi_1(x_1') \psi_1(x_1)d_p (\hat{x}_1,x_1)\\
&=
\sum_{x_1'} \psi_1(x_1') \min_{\hat{x}_1} \sum_{x_1} \psi_1(x_1)d_p (\hat{x}_1,x_1)\\
&=\min_{\hat{x}_1} \sum_{x_1} \psi_1(x_1)d_p (\hat{x}_1,x_1),
\end{aligned}
\end{equation}
which completes the proof.
\end{IEEEproof}

Then, with $f_1^{out}(x_1'|x_1,\emptyset)=\psi_1(x_1')$, we can have the optimal maximum privacy level as $\frac{1}{4}$.
\end{IEEEproof}

Then we obtain the maximum possible privacy gain for Proposition \ref{Prop: user 1 strategy}.
\end{IEEEproof}

\section{Proof of Proposition \ref{Prop:r_{in}}}
\label{Proof:Prop:r_{in}}
In Proposition \ref{Prop:r_{in}}, we want to optimize user 2's query strategy $r_{in}$ given that he has already found useful information in the cache, i.e., the service constraint is met. Then we need to discuss case-by-case according to the value of $x_1'$: $0 \leq x_1' \leq Q_2$, $Q_2 < x_1' \leq \frac{1}{2}$, $\frac{1}{2} < x_1' \leq 1-Q_2$, $1-Q_2 < x_1' \leq 1$.

\begin{figure*}[!t]
\centering
\includegraphics[width=5 in]{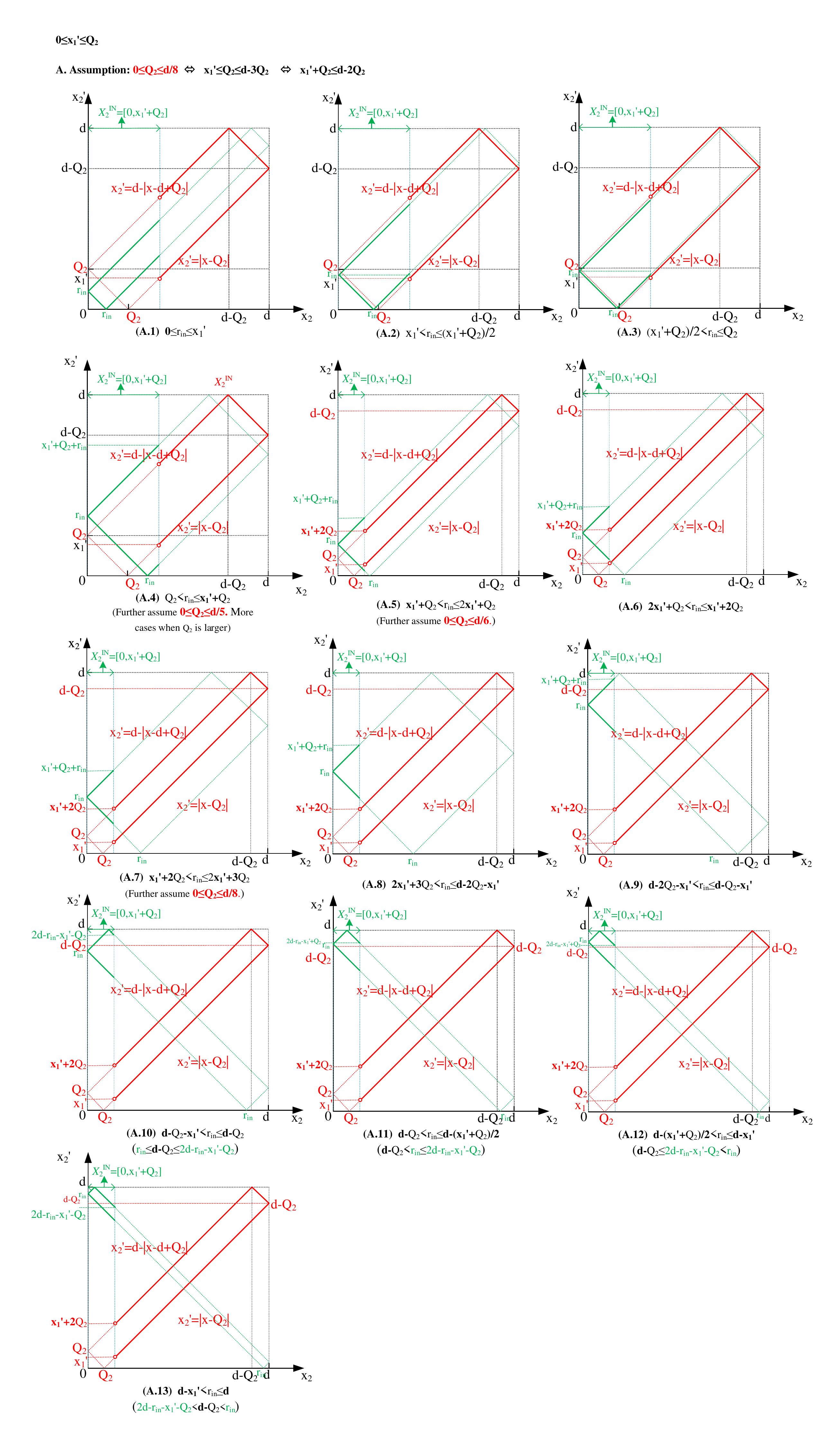}
\caption{User 2' query strategy under the discussion of $0 \leq x_1' \leq Q$. 
}
\label{Fig:2User-opt r_in-Case1}
\end{figure*}

Here we optimize Case 1 ($0 \leq x_1' \leq Q_2$) in detail and other cases are similar.
In Case 1, the optimization problem for $r_{in}$ is further listed in Fig.~\ref{Fig:2User-opt r_in-Case1}. 

\subsection{Detailed discussion of Case 1 for optimizing $r_{in}$}
\label{Proof:case1}

\textbf{(A.1)}
{\tiny
\begin{multline*}
2d \cdot \pi_2=
\int_{0}^{2r_{in}}|x-r_{in}|dx
+\int_{2r_{in}}^{x_1'+r_{in}}r_{in}dx+0
\\
+\int_{0}^{x_1'-r_{in}}r_{in}dx
+\int_{x_1'-r_{in}}^{x_1'+Q-2r_{in}}2r_{in}dx 
+\int_{x_1'+Q-2r_{in}}^{x_1'+Q} \frac{r_{in}+Q}{2}dx
\\
+\int_{x_1'+Q}^{x_1'+2Q-r_{in}}(Q-r_{in})dx
+\int_{x_1'+2Q-r_{in}}^{x_1'+2Q+r_{in}}\frac{r_{in}+Q}{2}dx \\
+\int_{x_1'+2Q+r_{in}}^{x_1'+3Q}0dx
+\int_{x_1'+3Q}^d Q dx
\\
+\int_{x_1'+Q}^{d-2Q} Q dx
+\int_{d-2Q}^d |x-(d-Q)| dx
\\
=2(x_1'+Q)r_{in}+8Q^2+2Q d-2Q x_1'.
\end{multline*}
}
To maximize $\pi_2$ for $r_{in} \in [0,x_1']$ in Case (A.1), we have $r_{in}=x_1'$.

\textbf{(A.2)}
{\tiny
\begin{multline*}
2d \cdot \pi_2=
\int_{0}^{r_{in}-x_1'}|2r_{in}-2x|dx
+\int_{r_{in}-x_1'}^{x_1'+r_{in}}|x-r_{in}|dx+0
\\
+\int_{0}^{x_1'+Q-2r_{in}}2r_{in}dx 
+\int_{x_1'+Q-2r_{in}}^{x_1'+Q} \frac{r_{in}+Q}{2}dx
\\
+\int_{x_1'+Q}^{x_1'+2Q-r_{in}}(Q-r_{in})dx
+\int_{x_1'+2Q-r_{in}}^{x_1'+2Q+r_{in}}\frac{r_{in}+Q}{2}dx
+\int_{x_1'+2Q+r_{in}}^{x_1'+3Q}0dx\\
+\int_{x_1'+3Q}^d Q dx
+\int_{x_1'+Q}^{d-2Q} Q dx
+\int_{d-2Q}^d |x-(d-Q)| dx
\\
=
2(x_1'+Q)r_{in}
+2Q d-2Q x_1'
-4Q^2.
\end{multline*}
}
To maximize $\pi_2$ for $r_{in} \in (x_1',\frac{x_1'+Q}{2}]$ in Case (A.2), we have $r_{in}=\frac{x_1'+Q}{2}$.

\textbf{(A.3)}
{\tiny
\begin{multline*}
2d \cdot \pi_2=
\int_{0}^{2r_{in}-x_1'-Q}|x-\frac{r_{in}+Q}{2}|dx
+\int_{2r_{in}-x_1'-Q}^{r_{in}-x_1'}|2r_{in}-2x|dx \\
+\int_{r_{in}-x_1'}^{x_1'+r_{in}}|x-r_{in}|dx+0
+\int_{0}^{x_1'+Q} \frac{r_{in}+Q}{2}dx
\\
+\int_{x_1'+Q}^{x_1'+2Q-r_{in}}(Q-r_{in})dx
+\int_{x_1'+2Q-r_{in}}^{x_1'+2Q+r_{in}}\frac{r_{in}+Q}{2}dx
+\int_{x_1'+2Q+r_{in}}^{x_1'+3Q}0dx \\
+\int_{x_1'+3Q}^d Q dx
+\int_{x_1'+Q}^{d-2Q} Q dx
+\int_{d-2Q}^d |x-(d-Q)| dx
\\
=
5r_{in}^2
-(\frac{9Q}{2}+\frac{7x_1'}{2})r_{in}+\frac{3}{2}Q x_1'+2Q d
-2Q^2
+\frac{3}{2}x_1^{'2}.
\end{multline*}
}
To maximize $\pi_2$ for $r_{in} \in (\frac{x_1'+Q}{2},Q]$ in Case (A.3), we have $r_{in}=Q$.

\textbf{(A.4)}
{\tiny
\begin{multline*}
2d \cdot \pi_2=
\int_{0}^{r_{in}-x_1'}|x-\frac{r_{in}+Q}{2}|dx
+0
+\int_{2r_{in}-x_1'-Q}^{x_1'+Q}|x-r_{in}|dx
\\
+\int_{0}^{x_1'+2Q-r_{in}} \frac{r_{in}+Q}{2}dx
+\int_{x_1'+2Q-r_{in}}^{x_1'+Q} (r_{in}-Q)dx
\\
+\int_{x_1'+Q}^{r_{in}+Q}|x-\frac{r_{in}+Q}{2}|dx
+\int_{r_{in}+Q}^{x_1'+3Q}\frac{r_{in}+Q}{2}dx
+\int_{x_1'+3Q}^{x_1'+2Q+r_{in}}2Qdx \\
+\int_{x_1'+2Q+r_{in}}^d Q dx
\\
+0
+\int_{x_1'+r_{in}}^{d-2Q} Q dx
+\int_{d-2Q}^d |x-(d-Q)| dx
\\
=
\frac{1}{2}r_{in}^2
+(\frac{1}{2}x_1'-\frac{3Q}{2})r_{in}
+2Q d-\frac{Q x_1'}{2}-Q^2.
\end{multline*}
}
To maximize $\pi_2$ for $r_{in} \in (Q,x_1'+Q]$ in Case (A.4), first we have that the maximum privacy is obtained when $r_{in}=Q$ or $r_{in}=x_1'+Q$. Then by comparing $\pi_2(r_{in}=Q)$ and $\pi_2(r_{in}=x_1'+Q)$, we have $r_{in}=x_1'+Q$.

\textbf{(A.5)}
{\tiny
\begin{multline*}
2d \cdot \pi_2=
\int_{0}^{r_{in}-x_1'}|x-\frac{r_{in}+Q}{2}|dx
+0
\\
+\int_{0}^{x_1'+2Q-r_{in}} \frac{r_{in}+Q}{2}dx
+\int_{x_1'+2Q-r_{in}}^{x_1'+Q} (r_{in}-Q)dx
\\
+\int_{x_1'+Q}^{r_{in}+Q}|x-\frac{r_{in}+Q}{2}|dx \\
+\int_{r_{in}+Q}^{x_1'+3Q}\frac{r_{in}+Q}{2}dx
+\int_{x_1'+3Q}^{x_1'+2Q+r_{in}}2Qdx
+\int_{x_1'+2Q+r_{in}}^d Q dx
\\
+0
+\int_{x_1'+r_{in}}^{d-2Q} Q dx
+\int_{d-2Q}^d |x-(d-Q)| dx
\\
=
-\frac{r_{in}^2}{2}
+(\frac{5}{2}x_1'+\frac{Q}{2})r_{in}
+2Q d-\frac{5}{2}Q x_1'-2Q^2-x_1^{'2}.
\end{multline*}
}
To maximize $\pi_2$ for $r_{in} \in (x_1'+Q,2x_1'+Q]$ in Case (A.5), we have $r_{in}=2x_1'+Q$.

\textbf{(A.6)}
{\tiny
\begin{multline*}
2d \cdot \pi_2=
\int_{0}^{x_1'+Q}|x-\frac{r_{in}+Q}{2}|dx
+0
\\
+\int_{0}^{x_1'+2Q-r_{in}} \frac{r_{in}+Q}{2}dx
+\int_{x_1'+2Q-r_{in}}^{x_1'+Q} (r_{in}-Q)dx
\\
+\int_{r_{in}-x_1'}^{r_{in}+Q}|x-\frac{r_{in}+Q}{2}|dx
+\int_{r_{in}+Q}^{x_1'+3Q}\frac{r_{in}+Q}{2}dx \\
+\int_{x_1'+3Q}^{x_1'+2Q+r_{in}}2Qdx
+\int_{x_1'+2Q+r_{in}}^d Q dx
\\
+0
+\int_{x_1'+r_{in}}^{d-2Q} Q dx
+\int_{d-2Q}^d |x-(d-Q)| dx
\\
=
\frac{r_{in}^2}{4}
+(x_1'-\frac{Q}{2})r_{in}
+2Q d-Q x_1'-\frac{7}{4}Q^2.
\end{multline*}
}
To maximize $\pi_2$ for $r_{in} \in (2x_1'+Q,x_1'+2Q]$ in Case (A.6), we have $r_{in}=x_1'+2Q$.

\textbf{(A.7)}
{\tiny
\begin{multline*}
2d \cdot \pi_2=
\int_{0}^{r_{in}-x_1'-2Q}|r_{in}-Q-2x|dx
+\int_{r_{in}-x_1'-2Q}^{x_1'+Q}|x-\frac{r_{in}+Q}{2}|dx
\\
+\int_{0}^{x_1'+Q} (r_{in}-Q)dx
\\
+\int_{r_{in}-x_1'}^{x_1'+3Q}|x-\frac{r_{in}+Q}{2}|dx
+\int_{x_1'+3Q}^{x_1'+2Q+r_{in}}2Qdx
+\int_{x_1'+2Q+r_{in}}^d Q dx
\\
+0
+\int_{x_1'+r_{in}}^{d-2Q} Q dx
+\int_{d-2Q}^d |x-(d-Q)| dx
\\
=
(Q+2x_1')r_{in}
+2Q d-3Q x_1'-4Q^2-x_1^{'2}.
\end{multline*}
}
To maximize $\pi_2$ for $r_{in} \in (x_1'+2Q,2x_1'+3Q]$ in Case (A.7), we have $r_{in}=2x_1'+3Q$.

\textbf{(A.8)}
{\tiny
\begin{multline*}
2d \cdot \pi_2=
\int_{0}^{x_1'+Q}|r_{in}-Q-2x|dx
\\
+\int_{0}^{x_1'+Q} (r_{in}-Q)dx
\\
+\int_{x_1'+3Q}^{r_{in}-x_1'}Qdx
+\int_{r_{in}-x_1'}^{x_1'+2Q+r_{in}}2Qdx
+\int_{x_1'+2Q+r_{in}}^d Q dx
\\
+\int_{x_1'+Q}^{r_{in}-x_1'-2Q}Qdx
+\int_{x_1'+r_{in}}^{d-2Q} Q dx
+\int_{d-2Q}^d |x-(d-Q)| dx
\\
=
(\frac{7}{2}Q+\frac{3}{2}x_1')r_{in}
+2Q d-\frac{17}{2}Q x_1'-13Q^2-\frac{1}{2}x_1^{'2}.
\end{multline*}
}
To maximize $\pi_2$ for $r_{in} \in (2x_1'+3Q,d-2Q-x_1']$ in Case (A.8), we have $r_{in}=d-2Q-x_1'$.

\textbf{(A.9)}
{\tiny
\begin{multline*}
2d \cdot \pi_2=
\int_{0}^{x_1'+Q}|r_{in}-Q-2x|dx
\\
+\int_{0}^{x_1'+Q} (r_{in}-Q)dx
\\
+\int_{x_1'+3Q}^{r_{in}-x_1'}Qdx
+\int_{r_{in}-x_1'}^{d}2Qdx
\\
+\int_{x_1'+Q}^{r_{in}-x_1'-2Q}Qdx
+0
+\int_{x_1'+r_{in}}^{2d-2Q-x_1'-r_{in}} |x-(d-Q)| dx
+\int_{2d-2Q-x_1'-r_{in}}^d |2d-2Q-2x| dx
\\
=
(\frac{3}{2}Q+\frac{3}{2}x_1')r_{in}
+2Q d-\frac{9}{2}Q x_1'-7Q^2-x_1^{'2}.
\end{multline*}
}
To maximize $\pi_2$ for $r_{in} \in (d-2Q-x_1',d-Q-x_1']$ in Case (A.9), we have $r_{in}=d-Q-x_1'$.

\textbf{(A.10)}
{\tiny
\begin{multline*}
2d \cdot \pi_2=
\int_{0}^{x_1'+Q}|r_{in}-Q-2x|dx
\\
+\int_{0}^{2d-2r_{in}-x_1'-Q} (r_{in}-Q)dx
+\int_{2d-2r_{in}-x_1'-Q}^{x_1'+Q}|x-\frac{2d-r+Q}{2}|dx
\\
+\int_{x_1'+3Q}^{r_{in}-x_1'}Qdx
+\int_{r_{in}-x_1'}^{d}2Qdx
\\
+\int_{x_1'+Q}^{r_{in}-x_1'-2Q}Qdx
+0
+\int_{2d-r_{in}-x_1'-2Q}^{r_{in}+x_1'} |x-\frac{2d-r+Q}{2}| dx
+\int_{r_{in}+x_1'}^d |2d-2Q-2x| dx
\\
=
- r_{in}^2
+(2d-\frac{Q+x_1'}{2})r_{in}
- 8Q^2 + 4Qd  - (13x_1'Q)/2 - d^2  + 2x_1'd   - \frac{3}{2}x_1^{'2}.
\end{multline*}
}
To maximize $\pi_2$ for $r_{in} \in (d-Q-x_1',d-Q]$ in Case (A.10), we have $r_{in}=d-Q'$.

\textbf{(A.11)}
Case (A.11) has the same total expected privacy gain as in Case (A.10), and we have $r_{in}=d-\frac{x_1'+Q'}{2}$.

\textbf{(A.12)} and \textbf{(A.13)}
{\tiny
\begin{multline*}
2d \cdot \pi_2=
\int_{0}^{2r_{in}-2d+x_1'-Q}|x-\frac{2d-r_{in}+Q}{2}|dx
+\int_{2r_{in}-2d+x_1'-Q}^{x_1'+Q}|r_{in}-Q-2x|dx
\\
+\int_{0}^{x_1'+Q}|x-\frac{2d-r+Q}{2}|dx
\\
+\int_{x_1'+3Q}^{r_{in}-x_1'}Qdx
+\int_{r_{in}-x_1'}^{d}2Qdx
\\
+\int_{x_1'+Q}^{r_{in}-x_1'-2Q}Qdx
+0
+\int_{2d-r_{in}-x_1'-2Q}^{r_{in}+x_1'} |x-\frac{2d-r+Q}{2}| dx
+\int_{r_{in}+x_1'}^d |2d-2Q-2x| dx
\\
=
- r_{in}^2
+(2d-Q)r_{in}
- 12Q^2 + 6Q  - 10x_1'Q - 1  + 2x_1'   - 2x_1^{'2}.
\end{multline*}
}
To maximize $\pi_2$ for $r_{in} \in (d-\frac{x_1'+Q}{2},d]$ in Case (A.12) and (A.13), we have
$r_{in}^*=d-\frac{Q}{2}$.
The maximum expected privacy gain is 
\begin{equation*}
\pi_2^*=
5Qd - 10Qx_1' + 2dx_1' - \frac{47}{4}Q^2 - \frac{9}{4} x_1^{'2}.
\end{equation*}

\section{Proof of Corollary \ref{Corol:Pi_2 appr reaches the min when x1'=d/2}}
\label{Proof:Corol:Pi_2 appr reaches the min when x1'=d/2}
\begin{IEEEproof}
According to Proposition \ref{Prop:r_{in}}, user 2's expected privacy in the proposed cooperative strategy with $Q_2 \leq \frac{1}{11}$ is given as
\begin{equation}
\label{Equ:pi_2 appr in proof}
\pi_2^{appr}(x_1')=
\begin{cases}
&\frac{5Q_2 - 10Q_2x_1' + 2x_1' - \frac{47}{4}Q_2^2 - \frac{9}{4} x_1^{'2}}{2},\\ &\text{if } 0 \leq x_1' \leq Q_2, \\
&\frac{1- 3Q_2^2 + 2Q_2 - 7 x_1' Q_2 - x_1' }{2},\\
&\text{if } Q_2 \leq x_1' \leq 1/2.
\end{cases}
\end{equation}
Notice that here we only discuss for $0 \leq x_1' \leq 1/2$ as the privacy function is symmetric on $x_1'$.
By minimizing the piecewise function of $\pi_2^{appr}(x_1')$, we obtain the minimum at $x_1'=1/2$ for $x_1' \in [0,1/2]$.
\end{IEEEproof}

\section{Proof of Corollary \ref{Corol:hide for user 2}}
\label{Proof:Corol:hide for user 2}
\begin{IEEEproof}
For user 2, following the simply-hiding strategy, user 2 will not query the LBS platform if $x_2 \in \mathcal{X}_2^{in}$. Take Fig. \ref{Fig:2User-hiding-Case1} for an example, only the query strategy in red line remains for $x_2 \in \mathcal{X}_2^{out}$. When $x_2 \in \mathcal{X}_2^{in}$, the adversary observes no reporting from user 2, and it can only randomly infer a location in $\mathcal{X}_2^{in}$.
\begin{figure}[!t]
\centering
\includegraphics[width=2 in]{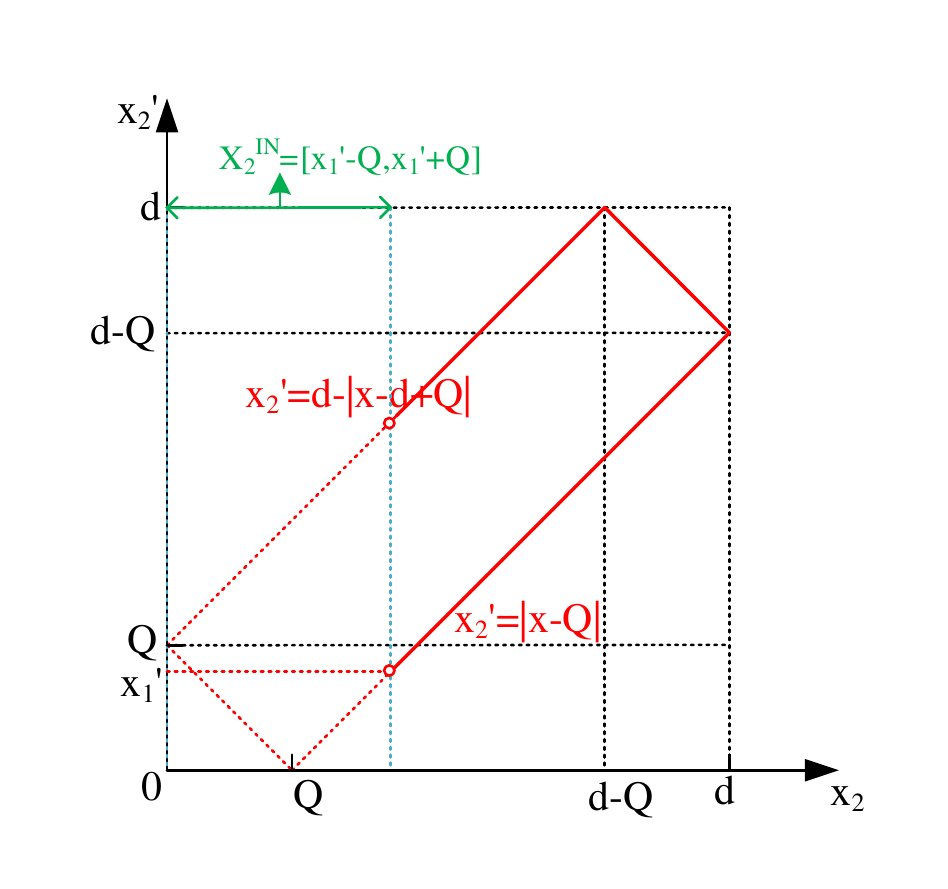}
\caption{User 2' query strategy under hiding-from-the-LBS scheme when $0 \leq x_1' \leq Q$.}
\label{Fig:2User-hiding-Case1}
\end{figure}

Similarly, by discussing for different $x_1'$, we obtain the hiding cooperative privacy for user 2 as
\begin{equation}
\label{Equ:pi_2 hide in proof}
\begin{aligned}
\pi_2^{hide}(x_1')=
\begin{cases}
&Q_2(1-2Q_2-\frac{x_1'}{2})+\frac{x_1'+Q_2}{4},\\
&\text{if }x_1' \in [0,Q_2), \\
&\frac{2Q_2 -5Q_2^2-2Q_2 x_1'}{2},\\
&\text{if }x_1' \in (Q_2,2Q_2], \\
&\frac{2Q_2+2Q_2 x_1'-Q_2^2-x_1^{'2}}{2},\\
&\text{if }x_1' \in (2Q_2,1/2].
\end{cases}
\end{aligned}
\end{equation}
Similarly as Appendix \ref{Proof:Corol:Pi_2 appr reaches the min when x1'=d/2}, here we only discuss for $x_1' \in [0,1/2]$ due to the symmetry of $x_1' \in [0,1]$.
Comparing the approximate cooperative privacy $\pi_2^{appr}$ in (\ref{Equ:pi_2 appr in proof}) and the hiding cooperative privacy $\pi_2^{hide}$ in (\ref{Equ:pi_2 hide in proof}) for user 2, we have $\pi_2^{appr}>\pi_2^{hide}$ for any $x_1' \in [0,1/2]$.
\end{IEEEproof}

\section{Proof of Lemma \ref{Lemma:X_k^{in}}}
\label{Proof:Lemma:X_k^{in}}
To prove Proposition \ref{Prop:infinite N leads to opt privacy}, Lemma \ref{Lemma:X_k^{in}} can be written as:
\begin{lemma}
\label{Lemma:X k {in}}
Former cooperative users tend to cover all the location set for latter users to meet the service constraints, i.e., $\mathcal{X}_N^{in} \overset{a.s.}{\to} [0,1]$. 
\end{lemma}
\begin{IEEEproof}
Consider that $k$ users have misreported $x_1',\cdots,x_k'$ contributing to a region $\mathcal{X}_k^{in}$ within which user $k+1$ can benefit from the former users.
As shown in Fig.~\ref{Fig:covered region}, the  region can be a continuous segment or disjoint continuous segments.
In both cases, we want to consider the user $i$'s worst-case reporting for increasing the whole range of $\mathcal{X}_i^{in}$, given any $|\mathcal{X}_k^{in}|<d$ (In this proof, we use a general $d$ instead of normalized $d=1$ to be clearer).
\begin{figure}[htb]
\centering
\includegraphics[width=2.5 in]{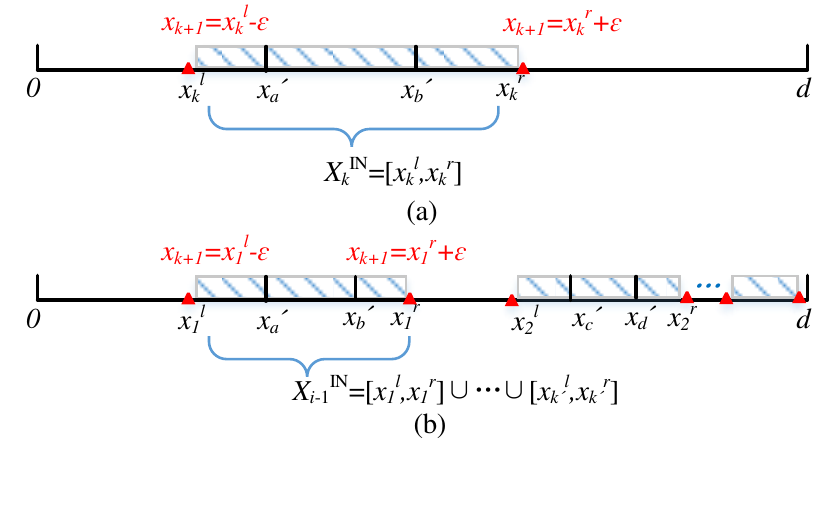}
\caption{For user $i$, the covered location set $\mathcal{X}_{i-1}^{in}$ might be (a) a continuous segment or (b) disjoint continuous segments.}
\label{Fig:covered region}
\end{figure}

\subsection{Continuous case}
Given the covered region $\mathcal{X}_{k}^{in}$ is a continuous segment as shown in Fig.~\ref{Fig:covered region}(a), we assume $\mathcal{X}_{k}^{in}=[x_k^l,x_k^r]$.
If user $k+1$'s real location $x_{k+1}$ locates outside the $\mathcal{X}_{k}^{in}$, i.e., $x_{k+1} \in \mathcal{X}_{k}^{out}$ with probability $1-\frac{|\mathcal{X}_{k}^{in}|}{d}$, the worst-case real location for enlarging the covered region is $x_{k+1}=x_k^l-\epsilon$ or $x_i=x_k^r+\epsilon$ ($\epsilon \to 0$).

Due to users' random reportings, as long as $x_k^l, x_k^r \in (Q,d-Q)$, we have $|\mathcal{X}_{k+1}^{in}|-|\mathcal{X}_{k}^{in}|\geq Q$ with probability $\frac{1}{2}\left( 1-\frac{|\mathcal{X}_{k}^{in}|}{d}\right) $. Then with a larger covered region, the continuous segment may become two disjoint segments (when $x_{k+1}'<x_k^l-Q$ or $x_{k+1}'>x_k^l+Q$). In the next case we will show that disjoint segments will converge into a continuous segment almost surely, hence here we only focus on how the continuous segment changes.

Notice that the probability 
$$\frac{1}{2}\left( 1-\frac{|\mathcal{X}_{k}^{in}|}{d}\right)>\frac{1}{2}\left( 1-\frac{d-2Q}{d}\right)=Q$$
always holds, given that $x_k^l, x_k^r \in (Q,d-Q)$. Then there exists a user $k_1<\infty$ such that
$x_{k_1}^l \in [0,Q]$, or $x_{k_1}^r \in [d-Q,Q]$ (as $|\mathcal{X}_{k_1}^{in}|=d-2Q$ is equivalent to $x_{k_1}^l=Q$ and $x_{k_1}^r=d-Q$).  
If $x_{k_1}^l \in [0,Q]$, the probability that user $k_1+1$ lies in $x_{k_1+1} \in [0,x_{k_1}^l) \subseteq \mathcal{X}_{k_1}^{out}$ is $\frac{x_{k_1}^l}{d}$, then we can have $x_{k_1+1}'\in [0,1]$ and $\mathcal{X}_{k_1+1}^{in}=[0,x_{k_1}^r]$ with probability $\frac{1}{2}\frac{x_{k_1}^l}{d}$.
\begin{lemma}
For all continuous segments $\mathcal{X}_{k_1}^{in}$ with $x_{k_1}^l \in [0,Q]$, there exists a user $k_2<\infty$ such that $\mathcal{X}_{k_2}^{in}=[0,x_{k_2}^r]$, or equivalently, $x_{k_2}^l=0$.
\end{lemma}

Given $\mathcal{X}_{k_2}^{in}=[0,x_{k_2}^r]$ and $x_{k_2}^r \in (Q,d-Q)$, similarly we have $|\mathcal{X}_{k_2+1}^{in}|-|\mathcal{X}_{k_2}^{in}|\geq Q$ with probability $\frac{1}{2}\left( 1-\frac{x_{k_2}^r}{d}\right) $, where 
$$\frac{1}{2}\left( 1-\frac{x_{k_2}^r}{d}\right) >\frac{1}{2}\left( 1-\frac{d-Q}{d}\right)=\frac{Q}{2}.$$
Thus we have that there exists a user $k_3<\infty$ such that
$x_{k_3}^r \in [d-Q,Q]$ (as $|\mathcal{X}_{k_3}^{in}|=d-Q$ is also equivalent to $x_{k_3}^r=d-Q$). Then similarly, there exist $k_4<\infty$ such that $x_{k_4}^r=d$, leading to $\mathcal{X}_{k_4}^{in}=[0,1]$.

If $x_{k_1}^r \in [d-Q,Q]$, the case is similar to $x_{k_1}^l \in [0,1]$.
Here we conclude that the continuous segment $\mathcal{X}_k^{in}$ converges to $[0,1]$ almost surely.

\subsection{Disjoint segments}
If the covered region $\mathcal{X}_k^{in}$ is composed of $k'$ disjoint segments, as shown in Fig.~\ref{Fig:covered region}(b), $\mathcal{X}_{k}^{in}$ can be denoted as 
\begin{align*}
\mathcal{X}_{k}^{in}&=\mathcal{X}_1 \cup  \mathcal{X}_2 \cup \cdots  \cup \mathcal{X}_{k'} \\
&=[x_1^l,x_1^r] \cup \cdots \cup [x_{k'}^l,x_{k'}^r], 
(\mathcal{X}_1 \cap  \mathcal{X}_2 \cap \cdots  \cap \mathcal{X}_{k'}=0).
\end{align*}

To show that the disjoint segments will always lead to a continuous segment, we focus on the OUT region among the disjoint segments, i.e., $\mathcal{X'}_k^{out}=\mathcal{X}_k^{out} \cap [x_1^l,x_{k'}^r]$.
If $x_{k+1} \in \mathcal{X}_k^{out} \cap [x_1^l,x_{k'}^r]$, there will be either one more segment or a larger segment with the reporting $x_{k+1}'$. Take two disjoint segments for example, $x_{k+1}'$ may add a new segment, increase the length of one segment, or merge two segments, as shown in Fig.~\ref{Fig:disjoint seg}.
Given $\mathcal{X'}_k^{out}=\mathcal{X}_k^{out} \cap [x_1^l,x_{k'}^r]$, the length $|\mathcal{X'}_k^{out}|< x_{k'}^r-x_1^l-2Q \cdot k'$, as one segment has a minimum length $2Q$. Then if $x_{k+1} \in \mathcal{X'}_k^{out}$ with probability $\frac{|\mathcal{X'}_k^{out}|}{d}$, there will be at most $\frac{x_{k'}^r-x_1^l-2Q \cdot k'}{2Q}$ more segments. That is, there is at most $k'_{\max}=k'+\frac{x_{k'}^r-x_1^l-2Q \cdot k'}{2Q}=\frac{x_{k'}^r-x_1^l}{2Q}$ segments after $\mathcal{X}_{k'}^{in}$.
\begin{figure}[htb]
\centering
\includegraphics[width=1.6 in]{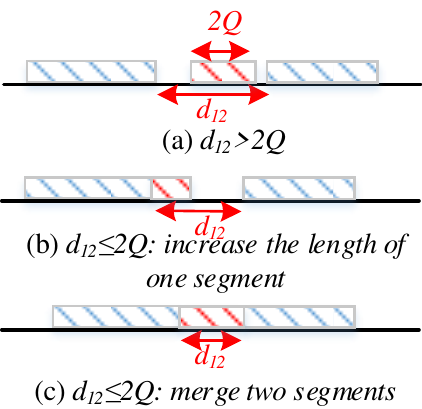}
\caption{Three cases of two disjoint segments: $x_{k+1}'$ may add a new segment, increase the length of one segment, or merge two segments.}
\label{Fig:disjoint seg}
\end{figure}

With $k'_{\max}$ disjoint segments, we denote the distance (length of the uncovered (out) region) between segment $a$ and $b$ as $d_{ab}<2Q$. Consider the worst-case when $x_{k+1}$ lies in the OUT region between segments $a$ and $b$, i.e., $x_{k+1} \in [x_a^r,x_b^l]$ with probability $\frac{d_{ab}}{d}$, we have $|\mathcal{X}_{k+1}^{in}|-|\mathcal{X}_{k}^{in}|\geq \frac{d_{ab}}{2}$  with probability $ \frac{1}{2}\frac{d_{ab}}{d}$ for each OUT region.
Hence, $d_{ab}$ converges to $0$ in distribution.
Recall that within the covered region $\mathcal{X}^{in}$, users' query strategy depends on $r_{in}$. Then there exists $k_5 < \infty$ such that after user $k_5$, $d_{ab}=x_b^l-x_a^r<r_{in}$ for all $a,b \in \{1,\cdots, k_{\max}'\}$. That is, all the OUT region is of short length and we want to cover the OUT region gap to the continuous segment.

Now consider the margin region of the cover region $\mathcal{X}^{in}$. For example, if $x_{k_5+1} \in [x_a^r-(r_{in}-d_{ab}),x_a^r]$ or $x_{k_5+1} \in [x_b^l,x_b^l+(r_{in}-d_{ab})]$ with probability $\frac{2(r_{in}-d_{ab})}{d}$, then by randomly reporting to the right/left side, the two segments $a$ and $b$ merge with probability $\frac{r_{in}-d_{ab}}{d}$, as shown in Fig.~\ref{Fig:disjoint seg}(c).
Hence all disjoint segments in $\mathcal{X'}_k^{out}$ merge with probability $1$ and we have that the case of disjoint segments will transfer to the case of a continuous segment we discussed earlier with probability $1$, and thus $X_{k}^{in}$ converges to $[0,1]$ almost surely, which concludes the proof.

Let $x_a'=\inf(\mathcal{X}_{i-1}^{in})+Q=x_k^l+Q$, $x_b'=\sup(\mathcal{X}_{i-1}^{in})-Q=x_k^r-Q$ denote the two most extreme reported data in set $\mathcal{X}_{k}^{in}$.
Given that user $i$ will take the strategy as the independent reporting in Proposition \ref{Prop: user 1 strategy}, i.e., his query location will be $x_i=x_a'-\epsilon$ or $x_i=x_b'+\epsilon$, leading to no enlargement of the region $\mathcal{X}_i^{in}$.
Otherwise, if $x_i$ locates at any other points in $[0,1]$, the minimum enlargement of the covered region $\mathcal{X}_i^{in}$ can be denoted by a positive size $\Delta$.

Then similarly, if $x_i=\inf(\mathcal{X}_j)-\epsilon$ or $x_i=\sup(\mathcal{X}_j)+\epsilon$ ($\epsilon \to 0$, $j=1,\cdots,I$), user $i$ will report a location $x_i'$ leading to no region enlargement. If $x_i$ locates at any other locations except for these points, we also denote the minimum enlargement of the covered region $\mathcal{X}_i^{in}$ by a positive size $\Delta$.
In both cases, we have 
\begin{align*}
&P(|\mathcal{X}_i^{in}|-|\mathcal{X}_{i-1}^{in}| \geq \Delta)\\
=&P(x_i \neq \inf(\mathcal{X}_j)-\epsilon \text{ and } x_i \neq \sup(\mathcal{X}_j)+\epsilon)
=1
\end{align*}
for all $i$.

Given the length of users' location region is $d$, let $k=\lceil \frac{d}{\Delta} \rceil <\infty$, then we have
\begin{equation*}
P(\mathcal{X}_k^{in}=[0,1])=1
\end{equation*}
and thus $
P(\lim_{N \to \infty} \mathcal{X}_N^{in}=[0,1])=1$.
\end{IEEEproof}

\section{Proof of Proposition \ref{Prop:infinite N leads to opt privacy}}
\label{Proof:Prop:infinite N leads to opt privacy}

Following the proof of Lemma \ref{Lemma:X_k^{in}} in Appendix \ref{Proof:Lemma:X_k^{in}}, we can proceed to prove Proposition \ref{Prop:infinite N leads to opt privacy}.

\begin{IEEEproof}
Recall Proposition \ref{Prop: user 1 strategy} that for an independent user (user 1), his maximum privacy gain is $\pi_1=\min(\frac{1}{4},Q_1-{Q_1^2})$, which equals $Q_1-{Q_1^2}$ under the assumption $Q_1 \leq 1/2$. Given Lemma \ref{Lemma:X k {in}}, for a user with order $i \geq k$, he can enjoy an expected privacy as
$\pi_i=\pi_k=\frac{1}{4}$.
For a user with order $1<i<k$, the privacy gain is $Q_1-{Q_1^2}<\pi_i<\frac{1}{4}$.

If there are totally $N$ users, consider a user with order $i$ with probability $\frac{1}{N}$, then he can achieve an expected privacy:
\begin{align*}
\Pi
&= 
\frac{1}{N}\sum_{i=1,\cdots,N}\pi_i
\\
&=\frac{1}{N}\left( Q_1-{Q_1^2}\right) +\frac{1}{N} \sum_{i=2,\cdots,k-1}\pi_i
+\frac{N-k+1}{N} \cdot \frac{1}{4}.
\end{align*}
As $N \rightarrow \infty$, we have $\lim_{N \rightarrow \infty} \Pi=\frac{1}{4}$, which shows the benefit of cooperative privacy protection.
\end{IEEEproof}

\section{Proof of Proposition \ref{Prop: performeance of appr}}
\label{Proof:Prop: performeance of appr}
\begin{IEEEproof}
Let $\pi_2^{opt}$ denote user 2's optimal cooperative privacy level obtained by Problem (\ref{Opt:user 2's reporting prob-cont}). Following the proof of Proposition \ref{Prop: user 1 strategy} in Appendix \ref{Proof:Prop: user 1 strategy}, we have the upper bound for a user's optimal privacy gain
\begin{equation*}
\pi_2^{opt} \leq \frac{1}{4}.
\end{equation*}

To explore how much privacy loss for user 2 by using the two randomized reportings in the binary approximate obfuscated query scheme, we have 
\begin{equation*}
\begin{aligned}
\max_{x_1'} \frac{\pi_2^{opt}}{\pi_2^{appr}(x_1')} \leq
&\frac{\frac{1}{4}}{\pi_2^{appr}(x_1'=\frac{1}{2})} \\
=&\frac{1}{1-Q^2-3Q} \leq 1.39.
\end{aligned} 
\end{equation*}
Given our assumption of a small $Q \leq \frac{1}{11}$, we have a bound $1.39$ for our approximation performance. This is equivalent to the statement of the performance loss in Proposition \ref{Prop: performeance of appr}. This shows the efficiency of our proposed cooperative privacy-preserving approach in this section.
\end{IEEEproof}

\section{Proof of Proposition \ref{Prop: order-2 small}}
\label{Proof:Prop: order-2 small}
\begin{IEEEproof}
Let user $i$ denote the first-order user and user $j$ denote the second-order user in the cooperation. Following the proof of Propositions \ref{Prop: user 1 strategy} and \ref{Prop:r_{in}}, the total privacy gain for the two users is given as
\begin{equation*}
\Pi(i,j)=Q_i-{Q_i^2}+\mathbb{E}_{x_i'}\pi_j^{appr}
\end{equation*}
with $\pi_j^{appr}$ in (\ref{Equ:pi_2 appr in proof}). Take the expectation over $x_i' \in [0,1]$, we have
\begin{align*}
\mathbb{E}_{x_i'}\pi_j^{appr}=&2\int_0^{Q_j} (\frac{5Q_j - 10Q_jx_i' + 2x_i' - \frac{47}{4}Q_j^2 - \frac{9}{4} x_i^{'2}}{2}) dx_i' \\
+&2\int_{Q_j}^{\frac{1}{2}}(\frac{1- 3Q_j^2 + 2Q_j - 7 x_i' Q_j - x_i' }{2})dx_i' \\
=&\frac{3}{2}Q_j^2-\frac{11}{4}Q_j^3-\frac{7}{8}Q_j-\frac{1}{8},
\end{align*}
which decreases with $Q_j$. Then we have
\begin{equation*}
\Pi(i,j)=Q_i-{Q_i^2}+3Q_j^2-11Q_j^3-\frac{7}{8}Q_j-\frac{1}{8}.
\end{equation*}

To compare $\Pi(1,2)$ and $\Pi(2,1)$, we have
\begin{align*}
&\Pi(1,2)-\Pi(2,1)\\
=&(11Q_1^3-\frac{25}{8}Q_1^2+Q_1)-(11Q_2^3-\frac{25}{8}Q_2^2+Q_2).
\end{align*}
Let $f(Q)=11Q^3-\frac{25}{8}Q^2+Q$, and we can prove that $f(Q)$ decreases with $Q$. Thus we have $f(Q_1) \geq f(Q_2)$ and $\Pi(1,2)\geq \Pi(2,1)$. The optimal sequence should be $\{1,2\}$.
\end{IEEEproof}

\section{Proof of Proposition \ref{Prop: order-1 small 1 large}}
\label{Proof:Prop: order-1 small 1 large}

\begin{IEEEproof}
Recall that in Proposition \ref{Prop: user 1 strategy}, we give the maximal privacy gain for an independent user $i$ as $\pi_i^{\max}=\min(Q_i-{Q_i^2},\frac{1}{4})$. Given $Q_2 \rightarrow \frac{1}{2}$, the privacy gain for user 2 approaches the maximum as $\pi_2 \rightarrow \frac{1}{4}$, no matter which order user 2 is at. Then we have
\begin{equation*}
\Pi(1,2)\rightarrow\pi_1+\frac{1}{4}
\end{equation*}
and
\begin{equation*}
\Pi(2,1)\rightarrow \frac{1}{4}+\mathbb{E}_{x_2'}\pi_1.
\end{equation*}
In this case, placing user 1 at the second order can preserve his privacy from the earlier query $x_2'$, thus we have the optimal sequence as $\{2,1\}$.
\end{IEEEproof}

\end{document}